\documentclass[10pt, twocolumn]{IEEEtran}

\newcommand{\Rmnum}[1]{\expandafter\@slowromancap\romannumeral #1@}
\makeatother
\usepackage{graphicx,cite,epsfig,amssymb,amsmath,multirow,extarrows}
\usepackage{graphicx,amsmath,amssymb,amsfonts,cite}
\usepackage{mathrsfs}
\usepackage{multirow}
\usepackage{algorithm}
\usepackage{graphicx}
\usepackage{subfigure}
\usepackage{epstopdf}
\usepackage{amsmath}
\usepackage{algorithmic}
\usepackage{float}
\usepackage{color}
\usepackage{bm}
\usepackage{stfloats}
\usepackage{xcolor}
\usepackage{ulem}
\usepackage{setspace}
\usepackage{diagbox}
\usepackage[square, comma, sort&compress, numbers]{natbib}

\usepackage{color}

\newtheorem{remark}{Remark}
\newtheorem{proof}{Proof}

\makeatother

\begin{document}
\bibliographystyle{ieeetr}
\title{\textcolor{black}{Downlink Precoding with Mixed Statistical and Imperfect Instantaneous CSI for Massive MIMO~Systems}}
\author{Shuang Qiu, Da Chen, Daiming Qu, Kai Luo, and Tao Jiang\\
\thanks{Manuscript received January 08, 2017; revised March 30, 2017 and July 22, 2017; accepted November 13, 2017. Date of publication ...; date of current version ... . This work was supported in part by the National Science Foundation for Distinguished Young Scholars of China (NSFC) with Grant numbers 61325004 and 61771216, the Fundamental Research Funds for the Central Universities, and National Science Foundation of China with Grant number 61701186.
The review of this paper was coordinated by Dr. Y.  Kawamoto. \emph{(corresponding Author: Tao Jiang.)}

Copyright (c) 2015 IEEE. Personal use of this material is permitted. However, permission to use this material for any other purposes must be obtained from the IEEE by sending a request to pubs-permissions@ieee.org.

S. Qiu, D. Chen, D.-M. Qu, K. Luo and T. Jiang are with the School of Electronic Information and Communications, Huazhong
University of Science and Technology, Wuhan 430074, P. R. China. (e-mails: sqiu, chenda, qudaiming, kluo@hust.edu.cn, tao.jiang@ieee.org).

}}
\maketitle

\begin{abstract}

In this paper, \textcolor{black}{the feasibility of a new downlink transmission mode in massive multi-input multi-output~(MIMO) systems is investigated} with two types of users, i.e., the users with only statistical channel state information (CSI) and the users with imperfect instantaneous CSI.
\textcolor{black}{The problem of downlink precoding design with mixed utilization of statistical and imperfect instantaneous CSI is addressed.}
We first theoretically analyze the impact of the mutual interference between the two types of users on their achievable rate.
Then, considering the mutual interference suppression, we propose an extended zero-forcing (eZF) and an extended maximum ratio transmission (eMRT) precoding methods to minimize the total transmit power of base station and to maximize the received signal power of users, respectively.
Thanks to the exploitation of statistical CSI, pilot-based channel estimation is avoided enabling more active users, higher system sum rate and shorter transmission delay.
\textcolor{black}{Finally, simulations are performed to validate the accuracy of the theoretical analysis and the advantages of the proposed precoding~methods.}

\end{abstract}
\section{Introduction}\label{Introduction}
Recently, massive multi-input multi-output (MIMO) has attracted much attention, where a base station (BS) is equipped with hundreds of antennas and simultaneously serves tens of users in the same time-frequency resource.
\textcolor{black}{Massive MIMO is a key enabler for the next generation communications due to its high energy efficiency and spectral efficiency~\cite{Ngo, Marzetta, Hoydis}.}

However, \textcolor{black}{the performance of massive MIMO systems highly depends on the accuracy of estimated instantaneous channel state information (CSI) and precoding methods \cite{JSDM, YHF, Kong2015, Saj2016}.}
Most of the existing precoding methods are based on accurate instantaneous CSI estimation, such as the classic zero-forcing~(ZF) and maximum ratio transmission (MRT) methods.
Unfortunately, it is a big challenge to acquire accurate instantaneous CSI in both frequency division duplexing~(FDD) and time division duplexing~(TDD) massive MIMO systems.
In FDD mode, the overheads for both downlink~pilot sequences and uplink channel feedback occupy lots of time-frequency resources~\cite{FDD1, FDD2, Apo2016} due to the large number of BS antennas, which makes FDD mode only support low-mobility and low-frequency scenarios~\cite{tenmyth}.
On the other hand, TDD mode is taken as an enabler for massive MIMO systems since downlink instantaneous CSI is obtained by estimating uplink CSI based on channel \textcolor{black}{reciprocity \cite{Shen2015}.} However, the downlink instantaneous CSI is not always accurate in practice due to calibration error in baseband-to-radio frequency chains \cite{ TDD2}.

\textcolor{black}{In contrast, statistical CSI such as channel covariance matrix is rather static and stable over fairly a long time, depending mainly on antenna parameters and surrounding environment.
The authors in~\cite{IViering2002} verified the long-term stability of statistical CSI by measurements
and the authors in~\cite{MestreNov.2008} stated that stable channel statistics could be easily obtained by long-term feedback or averaging over channel samples.
Therefore, statistical CSI is often assumed to be perfectly known by BSs at a much slower timescale with little or no feedback overhead~\cite{DaiJuly2016,HammarwallMarch2008}.
Due to the advantages of statistical CSI, it has drawn considerable interest and been widely used in massive MIMO systems.}
\textcolor{black}{In~\cite{LiuJan.152016, LiuSept.12014, LiuDec.15152016}, the authors proposed novel two-timescale hybrid analog/digital precoding schemes to reduce the number of required radio-frequency chains, where the analog precoder is only adaptive to statistical CSI.
By exploiting statistical CSI, such precoding schemes reduce channel estimation overhead and are robust to estimation latency.
Moreover, two-stage precoding methods were proposed in FDD massive MIMO systems to reduce downlink pilot and uplink feedback overheads, in which statistical CSI is utilized for user grouping and inter-group interference reduction~\cite{JSDM, Chen2014}.
Furthermore, statistical CSI is also adopted in hierarchical precoding structure to reduce inter-cell interference~\cite{LiuSept.152014, Jun2010,Emil2010}.}

It is seen that statistical CSI is taken as auxiliary information in most of the existing precoding schemes and effective instantaneous CSI of users is still required.
However, \textcolor{black}{instantaneous CSI of users might be unavailable at BSs, such as the users who have not been assigned pilots due to the deficiency of orthogonal pilot sequences and the currently inactive users who have not been contacting with BSs for a while.}
In this case, the BSs cannot exploit any array gains due to the absence of instantaneous CSI~\cite{tenmyth}.
Therefore, it is important to study how the BSs should act to contact with the users without instantaneous CSI.
Fortunately, statistical CSI can be easily obtained at BSs and be stable for a long time.
Thus, downlink transmission with only statistical CSI is a reasonable choice.
\textcolor{black}{Nevertheless, downlink precoding methods with only statistical CSI are rarely studied.} The optimal precoding scheme for single-user MIMO systems was studied in~\cite{single1, single2}.
Due to inter-user interference, researches on multiuser cases mainly focus on Rician fading channel in millimeter wave systems~\cite{Jin2015} and a simplified case with two single-antenna users and two antennas at BSs~\cite{Jin2012, VR2011}.
\textcolor{black}{Moreover,} the precoding methods in the two-user case require special orthogonality on the channel covariance matrices of users, which are difficult to be implemented in massive MIMO systems with tens of users.
\textcolor{black}{Recently, the authors in~\cite{ZhangJin2017} proposed a  statistical beamforming scheme with only statistical CSI based on the metric of signal-to-leakage-and-noise ratio maximization, which is applicable to massive MIMO systems.}

\textcolor{black}{In this paper,} to exploit the benefits of precoding with only statistical CSI and guarantee system performance, we consider a new downlink transmission mode where the BS simultaneously serves two types of users.
\textcolor{black}{The first type is named as type-S users who have not sent uplink pilots, such as the currently inactive users. The BS executes precoding only based on their statistical CSI.
The rest of users are named as type-C users and the BS obtains their imperfect instantaneous CSI by uplink pilot-aided channel estimation.}
Then, we investigate the feasibility of downlink transmission with mixed utilization of type-S users' statistical CSI and type-C users' instantaneous~CSI.

More specifically, our contributions are the following:
Firstly, a heuristic statistical beamforming method (SBM) is presented without considering mutual interference suppression between type-S and type-C users.
Then, the downlink achievable rate for each user is analyzed and approximate closed-form expression is derived.
\textcolor{black}{It is revealed that the mutual interference seriously impacts the performance of users.}
Hence, we propose two novel interference-suppressed precoding methods, i.e., the extended zero-forcing~(eZF) and the extended maximum ratio transmission~(eMRT) methods to minimize the total transmit power of BSs and to maximize the received signal power of users, respectively.
\textcolor{black}{
The proposed methods significantly suppress the mutual interference and improve the achievable rate of users.
\textcolor{black}{Moreover, the sum rate and sum spectral efficiency of massive MIMO systems are also improved compared to the systems with conventional precoding methods.}
Thanks to the use of statistical CSI, pilot-based channel estimation of type-S users is avoided making pilot overhead saved.
Then, the number of simultaneously served users is increased.
Simultaneously, the BS provides prompt and low-latency downlink transmission for users.
Finally, the accuracy of theoretical analysis and the advantages of the proposed methods are validated by simulation results.}

The rest of the paper is organized as follows. In Section~\ref{section:model}, the system and channel models are described.
\textcolor{black}{In Section \ref{section: Heuristic precoding}, the heuristic SBM precoding method is presented and the impact of mutual interference among users is analyzed.}
In Section~\ref{section: eZF precoding} and~\ref{section: precoding MRT}, the eZF and eMRT precoding methods are proposed, respectively. The simulation results are presented in Section~\ref{section:simu}. Conclusions are summarized in Section~\ref{section:conclusions}.

$\emph{Notations:}$ 
Boldface lowercase (uppercase) letters denote column vectors (matrices).
For matrix $\mathbf{A}$, notations $\mathbf{A}^*$, $\mathbf{A}^T$, $\mathbf{A}^H$, $\textrm{Tr}\left\{\mathbf{A}\right\}$, $\left\| \mathbf{A}\right\|_F$ and $\left\| \mathbf{A}\right\|$  indicate its conjugate, transpose, conjugate transpose, trace, Frobenius norm and 2-norm, respectively.
The maximum eigenvalue is denoted by~$\lambda_\mathrm{max}\left( \mathbf{A} \right)$ and the corresponding eigenvector is~$\mathbf{u}_\mathrm{max}\left( \mathbf{A} \right)$.
The pseudo inverse is~$\mathbf{A}^ \dag= \mathbf{A} \left( \mathbf{A}^H  \mathbf{A}\right)^{-1} $.
The notation $\mathbb{C}^{m \times n}$ represents a set of $m \times n$ matrices with complex entries and $\triangleq$ is used to denote a definition.
We use~${{\mathbf{I}}_{n}}$ to denote an $n\times n$ identity matrix.
The notation~$\mathbf{z}\sim \mathcal{CN}(0,\boldsymbol{\Sigma} )$ means~$\mathbf{z}$ is a complex Gaussian random vector with zero mean and covariance matrix $\boldsymbol{\Sigma} $.
\textcolor{black}{The expectation, variance and covariance operations of random variables are given as $\mathrm{E}\left\{ x \right\}$, $ \mathrm{Var}\left( x \right) = \mathrm{E}\left\{   \left| x- \mathrm{E}\left\{ x\right\} \right|^2 \right\}$ and~$ \mathrm{Cov}\left(x,y\right)= \mathrm{E}\left\{  xy \right\} -\mathrm{E}\left\{ x \right\}\mathrm{E}\left\{ y\right\} $,~respectively.}

\section{System and Channel Model }\label{section:model}

\textcolor{black}{This section describes the downlink transmission and channel models for the massive MIMO systems with the coexistence of type-C and type-S users.
In addition, the acquisition of statistical CSI of type-S users is discussed.}

\subsection{\textcolor{black}{Transmission Model}}

\begin{figure}[!h]
\centering
\includegraphics[width=3.0in]{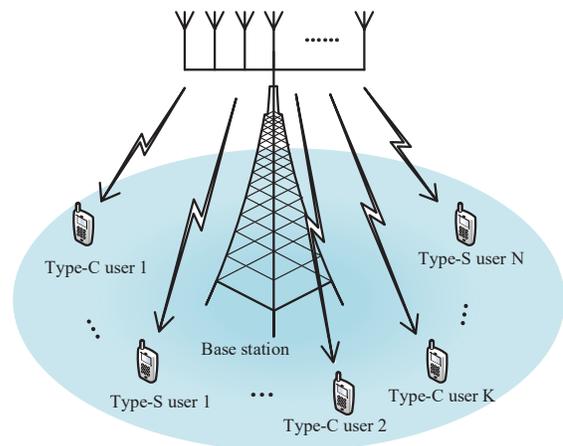}
\caption{ Downlink transmission with the coexistence of type-C and type-S users in a massive MIMO system.}
\label{system_model}
\end{figure}

As shown in Fig. \ref{system_model}, a BS with $M$ antennas simultaneously serves $K$ type-C users and $N$ type-S users in the downlink of a wideband massive MIMO system\footnote{Assume that type-C and type-S users have been scheduled by the BS. The specific strategy of user scheduling is beyond the scope of this paper and will be investigated in our future work. \textcolor{black}{Moreover, by using Orthogonal Frequency Division Multiplexing technology, the wideband channel is transformed into flat fading narrow band channels. Therefore, we focus on precoding design for a single sub-band and the precoders could be reused in other sub-bands.}}, where each user is equipped with a single antenna and $M \gg (K+N)$~\cite{Marzetta, Ngo, Hoydis}.
\textcolor{black}{We assume that the type-S users have strong channel correlation and their covariance matrices are rank deficient.}

The received signals ${y_{\mathrm{C},k}}$ and ${y_{\mathrm{S},n}}$ \textcolor{black}{by} the $k$-th type-C user and the $n$-th type-S user are expressed as
\begin{align}
 {y_{\mathrm{C},k}} & =  \mathbf{g}_{\mathrm{C},k}^{H}\left(\mathbf{W}_{\mathrm{C}} \mathbf{x}_{\mathrm{C}} + \mathbf{W}_{\mathrm{S}} \mathbf{x}_{\mathrm{S}}\right)+n_{\mathrm{C},k}, \label{eq 20170428.1}\\
 {y_{\mathrm{S},n}} & =  \mathbf{g}_{\mathrm{S},n}^{H} \left(\mathbf{W}_{\mathrm{C}} \mathbf{x}_{\mathrm{C}} + \mathbf{W}_{\mathrm{S}} \mathbf{x}_{\mathrm{S}}\right) +n_{\mathrm{S},n},\label{eq 20170428.2}
\end{align}
respectively, where $\mathbf{W}_\mathrm{C} \triangleq {\left[ {\begin{array}{*{20}{c}}
{{\mathbf{w}_{\mathrm{C},1}}}, & \cdots &, {{\mathbf{w}_{\mathrm{C},K}}}
\end{array}} \right]}\in \mathbb{C}^{M \times K} $ and $\mathbf{W}_\mathrm{S} \triangleq {\left[ {\begin{array}{*{20}{c}}
{{\mathbf{w}_{\mathrm{S},1}}}, & \cdots &, {{\mathbf{w}_{\mathrm{S},N}}}
\end{array}} \right]}\in \mathbb{C}^{M \times N} $ with ${\mathbf{w}_{\mathrm{C},k}} \in  \mathbb{C}^{M \times 1}$ and~${\mathbf{w}_{\mathrm{S},n}} \in  \mathbb{C}^{M \times 1}$ denoting the precoding vectors for the~$k$-th type-C and the $n$-th type-S users, respectively,
the vectors $\mathbf{x}_\mathrm{C} \triangleq {\left[ {\begin{array}{*{20}{c}}
{{ {x}_{\mathrm{C},1}}}, & \cdots &, {{ {x}_{\mathrm{C},K}}}
\end{array}} \right]^T}\in \mathbb{C}^{K \times 1} $ and $\mathbf{x}_\mathrm{S} \triangleq {\left[ {\begin{array}{*{20}{c}}
{{ {x}_{\mathrm{S},1}}}, & \cdots &, {{ {x}_{\mathrm{S},N}}}
\end{array}} \right]^T}\in \mathbb{C}^{N \times 1}$ are consisted of downlink data symbols which are independent, identically distributed~(i.i.d.) as ${x}_{\mathrm{C},k}, {x}_{\mathrm{S},n} \sim \mathcal{CN} (0,1)$,
the vectors~${\mathbf{g}_{\mathrm{C,}k} } \in \mathbb{C}^{M \times 1}$ and ${\mathbf{g}_{\mathrm{S,}n} } \in \mathbb{C}^{M \times 1}$ represent the channel vectors of the~$k$-th type-C and the $n$-th type-S users, respectively, the notations $n_{\mathrm{C},k}$ and $n_{\mathrm{S,}n}$ denote the additive white Gaussian noise (AWGN) distributed as $n_{\mathrm{C,k}}, n_{\mathrm{S,n}} \sim \mathcal{CN} (0,1)$.

We take the $k$-th type-C user as an example and its received signal~${y_{\mathrm{C},k}}$ is further expressed as
\vspace{-0.5em}
\begin{equation}
\begin{split}\label{eq 5}
 {y_{\mathrm{C},k}}  =&  \mathbf{g}_{\mathrm{C},k}^{H}{\mathbf{w}_{\mathrm{C},k}}{x_{\mathrm{C},k}}  + \sum\limits_{i = 1,i \ne k}^K   {\mathbf{g}_{\mathrm{C},k}^{H}{\mathbf{w}_{\mathrm{C},i}}{x_{\mathrm{C},i}}} \\
& + \sum\limits_{n = 1}^N \mathbf{g}_{\mathrm{C},k}^{H}{\mathbf{w}_{\mathrm{S},n}}{{ {x}}_{\mathrm{S},n}}  + {n_{\mathrm{C},k}},
\end{split}
\end{equation}
where the first term, $ \mathbf{g}_{\mathrm{C},k}^{H}{\mathbf{w}_{\mathrm{C},k}}{x_{\mathrm{C},k}} $, is the desired signal, the second term, $\sum\limits_{i = 1,i \ne k}^K {\mathbf{g}_{\mathrm{C},k}^{H}{\mathbf{w}_{\mathrm{C},i}}{x_{\mathrm{C},i}}}$, represents the interference from the other $K-1$ type-C users, and the third term, $\sum\limits_{n = 1}^N \mathbf{g}_{\mathrm{C},k}^{H}{\mathbf{w}_{\mathrm{S},n}}{{ {x}}_{\mathrm{S},n}} $, is the interference caused by the type-S users.
Then, based on the assumption of block fading channel model, the downlink ergodic achievable rate of the~$k$-th type-C and the $n$-th type-S user are respectively given as \cite{Shen2015}
\vspace{-0.5em}
\begin{align}
& {R_{\mathrm{C},k}}  = \nonumber \\
 &\mathrm{E}\left\{ {{{\hspace{-0.1em} \log }_2} \left( \hspace{-0.3em} {1 +\hspace{-0.3em} \frac{{|{\mathbf{g}_{\mathrm{C},k}^H}{\mathbf{w}_{\mathrm{C},k}}{|^2}}} {{\sum\limits_{{\substack{i = 1\\i \ne k}} }^K |{\mathbf{g}_{\mathrm{C},k}^H}{\mathbf{w}_{\mathrm{C},i}}{|^2}\hspace{-0.3em} + \hspace{-0.3em} \sum\limits_{n = 1}^N \hspace{-0.3em} |{\mathbf{g}_{\mathrm{C},k}^H}{\mathbf{w}_{\mathrm{S},n}}{|^2}\hspace{-0.3em} + \hspace{-0.3em}{1} }}   } \right)} \right\}, \label{eq 6} \\
&{R_{\mathrm{S},n}} = \nonumber \\
&  \mathrm{E}\left\{ {{{\log }_2}\left(\hspace{-0.3em} {1 +  \hspace{-0.3em} \frac{{|{\mathbf{g}_{\mathrm{S},n}^H}{\mathbf{w}_{\mathrm{S},n}}{|^2}}}{{\sum\limits_{{\substack{i = 1\\i \ne n}}}^N {|{\mathbf{g}_{\mathrm{S},n}^H}{\mathbf{w}_{\mathrm{S},i}}{|^2} \hspace{-0.3em}+ \sum\limits_{k = 1 }^K \hspace{-0.3em} |{\mathbf{g}_{\mathrm{S},n}^H}{\mathbf{w}_{\mathrm{C},k}}{|^2} + \hspace{-0.3em}{1}} }} } \right)} \right\}.\label{eq 20170503.1}
\end{align}

\subsection{\textcolor{black}{Channel Model}}

\textcolor{black}{Due to insufficient antenna spacing or finite local scatterers, the user channels are usually spatially correlated.
Therefore, we consider a correlated channel model in this paper and the channel vectors $\mathbf{g}_{\mathrm{C,}k}$ and $\mathbf{g}_{\mathrm{S,}n}$ are characterised as~\cite{Hoydis}
\begin{align} \label{eq 20170428.5}
\mathbf{g}_{\mathrm{C,}k} & =  \mathbf{\Phi}_{\mathrm{C,}k} ^{1/2} \mathbf{h}_{\mathrm{C,}k}, \\
\mathbf{g}_{\mathrm{S,}n} & =  \mathbf{\Phi}_{\mathrm{S,}n} ^{1/2} \mathbf{h}_{\mathrm{S,}n},
\end{align}
respectively, where  $\mathbf{\Phi}_{\mathrm{C,}k} \triangleq \mathrm{E}\left\{  \mathbf{g}_{\mathrm{C,}k} \mathbf{g}_{\mathrm{C,}k}^H \right\}$ and $\mathbf{\Phi}_{ \mathrm{S} ,n} \triangleq \mathrm{E}\left\{  \mathbf{g}_{ \mathrm{S} ,n} \mathbf{g}_{ \mathrm{S} ,n}^H \right\}$  are the channel covariance matrices, and~$\mathbf{h}_{\mathrm{C,}k}, \mathbf{h}_{\mathrm{S,}n}  \sim \mathcal{CN} \left(\mathbf{0}_{M \times 1},\mathbf{I}_M \right)$ are uncorrelated fast-fading channel vectors.
Note that this channel model is a general family of massive MIMO channels, which could also represent the i.i.d. Rayleigh case~(rich scattering environment) with~$\mathbf{\Phi}_{\mathrm{C,}k} =\mathbf{\Phi}_{\mathrm{S,}n}=\mathbf{I}_M$, for~$k=1,\dots,K, n=1,\dots,N.$}

\textcolor{black}{It is assumed that the BS holds the instantaneous CSI of type-C users by measurements on mutually orthogonal uplink pilot
sequences simultaneously transmitted by all the type-C users.
We ignore
the pilot contamination caused by the users with identical
pilot sequences in neighboring cells.
Then, the  minimum mean
square error (MMSE) estimate  $\widehat{\mathbf{{g}}}_{\mathrm{C,}k}$ of~$\mathbf{{g}}_{\mathrm{C,}k}$ is obtained as~\cite{YHF}
\begin{equation}\label{eq 20170429.1}
\hspace{-0.2em} \widehat{\mathbf{{g}}}_{\mathrm{C,}k} =\mathbf{ \Phi}_{\mathrm{C},k}\left(\hspace{-0.3em} \frac{1}{\tau p_u} \mathbf{I}_M +\mathbf{\mathbf{\Phi}}_{\mathrm{C},k} \right)^{-1} \hspace{-0.5em}\left( \mathbf{{g}}_{\mathrm{C,}k}+\hspace{-0.3em} \frac{1}{\sqrt{\tau p_u}} \mathbf{n}_{\mathrm{C,}k} \right),
\end{equation}
where $\tau$ is the length of pilot
sequences, the parameter $p_u$ is the pilot transmit power and
$\mathbf{n}_{\mathrm{C,}k} \sim  \mathcal{CN} \left( \mathbf{0}_{M \times 1},\mathbf{I}_M \right)$ is AWGN.
Then, the covariance matrix of $ \widehat{\mathbf{{g}}}_{\mathrm{C,}k}$ is given as
\begin{equation}\label{eq 20170429.2}
\hspace{-0.2em} \widehat{\mathbf{ \Phi}}_{\mathrm{C,}k} \hspace{-0.3em}  \triangleq \mathrm{E}\left\{ \widehat{\mathbf{{g}}}_{\mathrm{C,}k} \widehat{\mathbf{{g}}}_{\mathrm{C,}k}^H \right\} = \mathbf{\Phi}_{\mathrm{C},k} \hspace{-0.3em}  \left(\hspace{-0.2em}  \frac{1}{\tau p_u} \mathbf{I}_M +\mathbf{\Phi}_{\mathrm{C},k} \right)^{-1} \hspace{-0.5em} \mathbf{\Phi}_{\mathrm{C},k}.
\end{equation}
The instantaneous CSI of the $k$-th type-C user is expressed as
\begin{equation}\label{eq 20170429.3}
{\mathbf{{g}}}_{\mathrm{C,}k}= \widehat{\mathbf{{g}}}_{\mathrm{C,}k}+ \widetilde{{\mathbf{{g}}}}_{\mathrm{C,}k},
\end{equation}
where $\widetilde{{\mathbf{{g}}}}_{\mathrm{C,}k} \sim \mathcal{CN} \left(\mathbf{0}_{M \times 1}, \mathbf{\Delta}_{\mathrm{C,}k} \right)$ is estimation error with~$ \mathbf{\Delta}_{\mathrm{C,}k}= \mathbf{\Phi}_{\mathrm{C},k}-  \widehat{\mathbf{ \Phi}}_{\mathrm{C,}k}$.
Moreover, the estimation error~$\widetilde{{\mathbf{{g}}}}_{\mathrm{C,}k}$ is uncorrelated with the estimated channel $\widehat{\mathbf{{g}}}_{\mathrm{C,}k}$ due to the orthogonal properties of MMSE receiver.}

\subsection{\textcolor{black}{Acquisition of Statistical CSI}}

\textcolor{black}{We assume that the BS only holds the statistical CSI of type-S users in this paper.
The rationality of this assumption is twofold.
First, the instantaneous CSI of users is sometimes unavailable at the BS, such as the inactive users who have not sent uplink pilots.
Secondly, the variation of statistical CSI is roughly two orders of magnitude slower than instantaneous CSI~\cite{IViering2002}, which makes it much easier to obtain an accurate estimation of statistical CSI, i.e., covariance~matrices.}

\textcolor{black}{Covariance matrix estimation is an important issue in practice.
Conventional  methods are to average over the sample covariance matrices constructed from observations~\cite{MestreNov.2008}.
Recently, more sophisticated algorithms were proposed to reduce the overhead of channel covariance estimation.
The authors in~\cite{LiuJune2015} proposed a rank-deficient oracle approximating shrinkage~estimator, which significantly reduces the demand of independent channel samples, especially for
highly correlated channels.
Besides, a two-stage estimator was also proposed to improve the accuracy of covariance matrix estimation~\cite{BjoernsonNov.2016}.}
\textcolor{black}{Moreover, the downlink channel covariance matrices in TDD mode are able to be directly obtained via uplink training and channel reciprocity~\cite{ Ngo}.
Another feasible method is to average over the past channel measurements recorded by the BS.
In this way, channel covariance matrices are obtained via the pilots already used for uplink channel estimation and no extra overhead is needed for the estimation of channel covariance~matrices.}
%
%
%

\textcolor{black}{In this paper, we assume that the BS knows the channel statistics of type-S users via long-term transmission.
Then, the BS schedules the type-S users with rank-deficient covariance matrices and directly utilizes the covariance matrices to conduct downlink precoding.
}
\textcolor{black}{Note that, when instantaneous CSI of type-S users is unknown to the BS, natural questions are how to  precode for type-C and type-S users, and how mutual interference influences their achievable rate.
Next, the impact of the mutual interference will be analyzed in Section~\ref{section: Heuristic precoding}.}

\section{Impact of Mutual Interference on Achievable~Rate}\label{section: Heuristic precoding}
\textcolor{black}{In this section, the SBM method is first presented as a heuristic method without considering the mutual interference suppression between type-C and type-S users.
Then, the closed-form expressions for the achievable rate of users are derived and the impact of mutual interference on their achievable rate is analyzed.}

\subsection{ The Statistical Beamforming Method (SBM) }
Inspired by the statistical beamforming methods proposed in~\cite{Jin2012, VR2011} for a two-user transmission case, we design the statistical beamforming vector for the $n$-th type-S user as
\begin{equation} \label{heuheu9}
\mathbf{w}_{\mathrm{S},n}^\mathrm{SBM}=\sqrt{p_d } \mathbf{u}_\mathrm{max} \left ( \mathbf{\Phi}_{\mathrm{S},n} \right),
\end{equation}
where $p_d$ is the downlink transmit power to each user.
Moreover,
\textcolor{black}{the precoding vector for the $k$-th type-C user adopts the classic MRT method}\footnote{ \textcolor{black}{The main purpose of this section is to explore the impact of mutual interference between type-C and type-S users. Precoding methods for type-C users mainly affect the interference among type-C users. Thus, we only consider the representative MRT precoding due to the space limitation.}} \textcolor{black}{given~as}
\begin{equation}\label{heu c 1}
\mathbf{w}_{\mathrm{C},k}^\mathrm{SBM}=\sqrt{p_d}  \frac{\widehat{\mathbf{g}}_{{\mathrm{C},k}}}{\left\| {\widehat{\mathbf{g}}_{{\mathrm{C},k}}} \right\|}.
\end{equation}

Next, we will derive the closed-form expressions of the ergodic achievable rate for the $k$-th~type-C and $n$-th type-S users in (\ref{eq 6}) and (\ref{eq 20170503.1}) to analyze the impact of mutual interference between the two types of users.
However, the closed-form expressions are difficult to be obtained due to the unknown instantaneous CSI of type-S users and the correlation between the numerator and denominator of (\ref{eq 6}) and (\ref{eq 20170503.1}).
\textcolor{black}{Therefore, we introduce the following lemma to overcome the difficulties.}

\textcolor{black}{\emph{Lemma 1 {(}\cite[ {Lemma} 1]{ZQi} {)}:}
When $X \triangleq \sum\limits_{i = 1}^{{t_1}} {{X_i}}$ and $Y \triangleq \sum\limits_{j = 1}^{{t_2}} {{Y_j}}$ are both non-negative, we have
\begin{equation} \label{eq3 6}
 \mathrm{E} \left\{ \log_2 \left( 1+\frac{X}{Y} \right) \right\} \rightarrow  \log_2 \left( 1+\frac{\mathrm{E} \left\{ X \right\}}{\mathrm{E} \left\{ Y \right\}}  \right), \; \textrm{as} \; t_1, t_2 \rightarrow \infty.
\end{equation}
\begin{remark}
The above lemma does not require the independence between random variables $X$ and $Y$.
Moreover, the authors in~\cite{ZQi} also stated that (\ref{eq3 6}) could be utilized to obtain approximate closed-form expression of the ergodic achievable rate in massive MIMO systems due to the large number of BS antennas.
We also verify the accuracy in the Section~\ref{section:simu}.
\end{remark}}

\subsection{\textcolor{black}{Impact Analysis of Mutual Interference on the Achievable Rate for Type-C users}} \label{subsection rate C}

Based on (\ref{eq 6}) and \emph{Lemma} 1, an approximate closed-form expression ${\widetilde{R}_{\mathrm{C},k}} $ of the achievable rate for the $k$-th type-C user could be expressed as
\begin{equation}\label{eqeq 1}
{\widetilde{R}_{\mathrm{C},k}} = \log_2 \Bigg(  1+  \frac{S_{\mathrm{C},k}}{I_{\mathrm{C1},k}+I_{\mathrm{C2},k}+ {1}} \Bigg),
\end{equation}
where $S_{\mathrm{C},k}$, $I_{\mathrm{C1},k}$ and $I_{\mathrm{C2},k}$ are given by
\begin{align}
S_{\mathrm{C},k} & \triangleq \mathrm{E}\left\{ {\left|{\mathbf{g}_{\mathrm{C},k}^H}{\mathbf{w}_{\mathrm{C},k}} \right|^2}  \right\}, \label{a1}\\
I_{\mathrm{C1},k}& \triangleq \sum\limits_{i = 1,i \ne k}^K \mathrm{E}\left\{{|{\mathbf{g}_{\mathrm{C},k}^H}{\mathbf{w}_{\mathrm{C},i}}{|^2} }  \right\},\label{a2}\\
I_{\mathrm{C2},k}& \triangleq\sum\limits_{n = 1}^N \mathrm{E}\left\{ |{\mathbf{g}_{\mathrm{C},k}^H}{\mathbf{w}_{\mathrm{S},n}}{|^2} \right\},\label{a3}
\end{align}
respectively. The symbol $S_{\mathrm{C},k}  $ in (\ref{a1}) is the average signal power of the~$k$-th type-C user,~$I_{\mathrm{C1},k}$ in~(\ref{a2}) is the average interference power from the other $K-1$ type-C users and~$I_{\mathrm{C2},k}$ in~(\ref{a3}) is the average interference power from type-S users.

Then, the approximate closed-form expression $\widetilde{R}_{\mathrm{C},k}$ for the ergodic achievable rate of the~$k$-th type-C user with the SBM method could be obtained as~(\ref{heuheu22}), shown at the top of next page, where~$ \mathbf{\mathbf{F}}\left(\mathbf{A},\mathbf{B}\right) = \mathrm{ tr}\left( \mathbf{A} \right) \mathrm{ tr}\left( \mathbf{B} \right)+\mathrm{ tr}\left( \mathbf{AB} \right)-\mathrm{ tr}\left( \mathbf{D}\left(\mathbf{A}\right) \mathbf{D}\left(\mathbf{B}\right) \right)$ with $\mathbf{D}\left( \mathbf{A} \right)$ denoting a diagonal matrix consisting of the diagonal elements of matrix $\mathbf{A}$.
The proof of~(\ref{heuheu22}) is given in Appendix \ref{app_rate_C}.
It is shown from (\ref{heuheu22}) that the ergodic achievable rate of type-C users is influenced by the spatial correlation between type-C and type-S users.
Once the two types of users have identical dominant eigenmodes, type-C users will suffer serious interference from type-S~users.
\newcounter{TempEqCnt}
\setcounter{TempEqCnt}{\value{equation}}
\setcounter{equation}{17}
\begin{figure*}[ht]
\textcolor{black}{
\begin{equation} \label{heuheu22}
\begin{split}
&\hspace{-0.9em}{\widetilde{R}_{\mathrm{C},k}}= \\
&\hspace{-0.9em}\log_2 \hspace{-0.3em} \left( \hspace{-0.3em} 1\hspace{-0.3em} + \hspace{-0.3em} \frac{ \mathrm{tr} \left(\widehat{\mathbf{ \Phi}}_{\mathrm{C},k}  \right) + \frac{\mathrm{tr} \left(\widehat{\mathbf{ \Phi}}_{\mathrm{C},k} \mathbf{\Delta}_{\mathrm{C},k} \right)}{\mathrm{tr} \left(\widehat{\mathbf{ \Phi}}_{\mathrm{C},k}  \right)} - \frac{ \mathbf{\mathbf{F}}\left( \widehat{\mathbf{ \Phi}}_{\mathrm{C},k}^{1/2} \mathbf{\Delta}_{\mathrm{C},k} \widehat{\mathbf{ \Phi}}_{\mathrm{C},k}^{1/2} , \widehat{\mathbf{ \Phi}}_{\mathrm{C},k} \right)   }{\mathrm{tr} ^2\left(\widehat{\mathbf{ \Phi}}_{\mathrm{C},k}  \right)}  + \frac{ \mathbf{ F}\left( \widehat{\mathbf{ \Phi}}_{\mathrm{C},k},\widehat{\mathbf{ \Phi}}_{\mathrm{C},k} \right)    \mathrm{tr} \left(\widehat{\mathbf{ \Phi}}_{\mathrm{C},k} \mathbf{\Delta}_{\mathrm{C},k} \right)}{\mathrm{tr} ^3\left(\widehat{\mathbf{ \Phi}}_{\mathrm{C},k}  \right)}}
 { \hspace{-0.3em} \sum\limits_{{\substack{i=1\\i\neq k}}}^K \hspace{-0.3em} \left(\hspace{-0.3em} \frac{\mathrm{tr} \left(\widehat{\mathbf{ \Phi}}_{\mathrm{C},i} \mathbf{\Phi}_{\mathrm{C},k} \right)}{\mathrm{tr} \left(\widehat{\mathbf{ \Phi}}_{\mathrm{C},i}  \right)} - \hspace{-0.3em} \frac{ \mathbf{\mathbf{F}}\left( \widehat{\mathbf{ \Phi}}_{\mathrm{C},k}^{1/2} \mathbf{\Phi}_{\mathrm{C},k} \widehat{\mathbf{ \Phi}}_{\mathrm{C},i}^{1/2} , \widehat{\mathbf{ \Phi}}_{\mathrm{C},i} \right)   }{\mathrm{tr} ^2\left(\widehat{\mathbf{ \Phi}}_{\mathrm{C},i}  \right)} +\hspace{-0.3em} \frac{ \mathbf{ F}\left( \widehat{\mathbf{ \Phi}}_{\mathrm{C},i},\widehat{\mathbf{ \Phi}}_{\mathrm{C},i} \right)    \mathrm{tr} \left(\widehat{\mathbf{ \Phi}}_{\mathrm{C},i} \mathbf{\Phi}_{\mathrm{C},k} \right)}{\mathrm{tr} ^3\left(\widehat{\mathbf{ \Phi}}_{\mathrm{C},i}  \right)} \right)+\hspace{-0.3em} \sum\limits_{n = 1}^N \hspace{-0.3em} \mathbf{u}^H_\mathrm{max} \left ( \mathbf{\Phi}_{\mathrm{S},n} \right)  \mathbf{\Phi} _{{\mathrm{C},k}}  \mathbf{u}_\mathrm{max} \left ( \mathbf{\Phi}_{\mathrm{S},n} \right) +\hspace{-0.3em} \frac{1}{p_d}} \right)
\end{split}
\end{equation}
\hrulefill}
\end{figure*}
\setcounter{equation}{\value{TempEqCnt}}

\newcounter{TempEqCnt1}
\setcounter{TempEqCnt1}{\value{equation}}
\setcounter{equation}{23}
\begin{figure*}[ht]
\textcolor{black}{\begin{align} \label{heuheu23}
&\hspace{-0.9em}  {\widetilde{R}_{\mathrm{S},n}} = \nonumber \\
&\hspace{-0.9em}  \log_2 \hspace{-0.3em}  \left( \hspace{-0.3em} 1+ \hspace{-0.3em} \frac{  \lambda_\mathrm{max} \left ( \mathbf{\Phi}_{\mathrm{S},n} \right)}{\hspace{-0.3em} \sum\limits_{{\substack{j=1\\j\neq n}}}^N \hspace{-0.3em} \mathbf{u}_\mathrm{max}^H \left ( \mathbf{\Phi}_{\mathrm{S},j} \right)  \mathbf{\Phi}_{\mathrm{S},n}   \mathbf{u}_\mathrm{max} \left ( \mathbf{\Phi}_{\mathrm{S},j} \right) \hspace{-0.3em} + \hspace{-0.5em} \sum\limits_{k = 1 }^K \hspace{-0.3em} \left( \hspace{-0.3em} \frac{\mathrm{tr} \left(\widehat{\mathbf{ \Phi}}_{\mathrm{C},k} \mathbf{\Phi}_{\mathrm{S},n} \right)}{\mathrm{tr} \left(\widehat{\mathbf{ \Phi}}_{\mathrm{C},k}  \right)} \hspace{-0.3em} -\hspace{-0.3em} \frac{ \mathbf{\mathbf{F}}\left( \widehat{\mathbf{ \Phi}}_{\mathrm{C},k}^{1/2} \mathbf{\Phi}_{\mathrm{S},n} \widehat{\mathbf{ \Phi}}_{\mathrm{C},k}^{1/2} , \widehat{\mathbf{ \Phi}}_{\mathrm{C},k} \right)   }{\mathrm{tr} ^2\left(\widehat{\mathbf{ \Phi}}_{\mathrm{C},k}  \right)}  +\hspace{-0.3em} \frac{ \mathbf{ F}\left( \widehat{\mathbf{ \Phi}}_{\mathrm{C},k},\widehat{\mathbf{ \Phi}}_{\mathrm{C},k} \right)    \mathrm{tr} \left(\widehat{\mathbf{ \Phi}}_{\mathrm{C},k} \mathbf{\Phi}_{\mathrm{S},n} \right)}{\mathrm{tr} ^3\left(\widehat{\mathbf{ \Phi}}_{\mathrm{C},k}  \right)} \right)+\hspace{-0.3em} \frac{1}{p_d}}   \right)
\end{align}
\hrulefill}
\end{figure*}
\setcounter{equation}{\value{TempEqCnt1}}

To intuitively present the received signal and interference power, we consider a simpler case where all users have i.i.d. Rayleigh fading~channels.
Then, we have $\mathbf{\Phi}_{\mathrm{C},k}=\mathbf{\Phi}_{\mathrm{S},n}= \mathbf{I}_M $, $\widehat{\mathbf{ \Phi}}_{\mathrm{C},k}=\frac{\tau p_u}{\tau p_u +1}\mathbf{I}_M$ and $\mathbf{\Delta}_{\mathrm{C},k}=\frac{1}{\tau p_u +1}\mathbf{I}_M$ for $ \forall k$ and $\forall n$.
Substituting these items into (\ref{heuheu22}), we obtain the approximate closed-form expression for the ergodic achievable rate under i.i.d. Rayleigh channels~as
\setcounter{equation}{19}
\begin{equation}\label{eq 20170514.1}
  {\widetilde{R}^{\mathrm{i.i.d}}_{\mathrm{C},k}}=\log_2 \left( 1+ \frac{\tau p_u M +1}{\left(\tau p_u+1 \right) \left(K-1 + N  + \frac{1}{p_d}\right)} \right).
\end{equation}
\textcolor{black}{It is seen from (\ref{eq 20170514.1}) that
the average received signal power of type-C users depends on the number of BS antennas, length of pilot sequences and pilot transmit power.
In i.i.d. Rayleigh fading case, there is no spatial correlation between the channels of type-C and type-S users.
The type-S users cause the same degree of interference to the $k$-th type-C user as the other $K-1$ type-C users.
However, once there exits spatial correlation, the $k$-th type-C user will suffer more interference from type-S users.
Therefore, it is necessary to consider the mutual~interference suppression between type-C and type-S~users.}

\subsection{\textcolor{black}{Impact Analysis of Mutual Interference on the Achievable Rate for Type-S Users}}\label{subsection: 3.c}
\textcolor{black}{The impact of interference from type-C users on type-S users is analyzed} in this subsection. Based on \emph{Lemma} 1, an approximate achievable rate expression ${\widetilde{R}_{\mathrm{S},n}}$ of the~$n$-th type-S user is given~as
\begin{equation}\label{a4}
{\widetilde{R}_{\mathrm{S},n}} = \log_2 \left(  1+   \frac{{S_{\mathrm{S},n}}}{{I_{\mathrm{S1},n}}+{I_{\mathrm{S2},n}}+ 1} \right),
\end{equation}
where ${S_{\mathrm{S},n}} $, ${I_{\mathrm{S1},n}}$ and ${I_{\mathrm{S2},n}}$ are given by
\begin{align}
{S_{\mathrm{S},n}} & \triangleq \mathrm{E}\left\{ {\left|{\mathbf{g}_{\mathrm{S},n}^H}{\mathbf{w}_{\mathrm{S},n}} \right|^2}  \right\}, \label{a5} \\
{I_{\mathrm{S1},n}} & \triangleq \sum\limits_{j = 1,j \ne n}^N \mathrm{E}\left\{{ \left|{\mathbf{g}_{\mathrm{S},n}^H}{\mathbf{w}_{\mathrm{S},j}} \right|^2 }  \right\} , \label{a6}\\
{I_{\mathrm{S2},n}} & \triangleq \sum\limits_{k=1}^K \mathrm{E}\left\{ \left\| {\mathbf{g}_{\mathrm{S},n}^H}{\mathbf{w}_{\mathrm{C},k}}\right\|^2  \right\},\label{a7}
\end{align}
respectively, where ${S_{\mathrm{S},n}}$ in (\ref{a5}) denotes the average received signal power,
the symbol ${I_{\mathrm{S1},n}}$ in~(\ref{a6}) is the average interference power from the other~$N-1$ type-S users to the~$n$-th type-S user and ${I_{\mathrm{S2},n}}$ in (\ref{a7}) is the average interference power from the $K$ type-C users.
By following a similar proof to (\ref{heuheu22}), the approximate closed-form expression $\widetilde{R}_{\mathrm{S},n}$ of the ergodic achievable rate for the $n$-th type-S user is given in (\ref{heuheu23}), shown at the top of this~page.


When all the users have i.i.d. Rayleigh fading
channels, a simplified approximate closed-form expression for the ergodic achievable rate of the~$n$-th type-S user could be obtained as
\setcounter{equation}{25}
\begin{equation}\label{eq 20170514.2}
  \widetilde{R}^{\mathrm{i.i.d.}}_{\mathrm{S},n}=\log_2 \left(1+\frac{1}{N-1 + K+\frac{1}{p_d} }  \right).
\end{equation}
Combining with  (\ref{heuheu23})
and (\ref{eq 20170514.2}), we could conclude that the received signal power of type-S users highly depends on their channel statistics.
When the type-S users have stronger channel correlation, the received signal power becomes larger.
On the contrary, when the $n$-th type-S user has i.i.d. Rayleigh channel, the received signal power is only proportional to~$\lambda_\mathrm{max} \left ( \mathbf{\Phi}_{\mathrm{S},n} \right)=1$.
Hence, the BS tends to take the users with strong channel correlation as type-S users in~practice.

\textcolor{black}{Moreover, it is found from (\ref{heuheu23}) that the interference among type-S users could be eliminated by proper scheduling, such as selecting the type-S users with non-overlapping  angle-of-arrivals (AOAs)~\cite{YHF}.
Thus, we~have
 \begin{align}
 \mathbf{u}_\mathrm{max}^H \left ( \mathbf{\Phi}_{\mathrm{S},j} \right)  \mathbf{\Phi}_{\mathrm{S},n}   \mathbf{u}_\mathrm{max} \left ( \mathbf{\Phi}_{\mathrm{S},j} \right) &= 0,  \; \forall {{j}} \ne n.
\end{align}
Note that, the idea of scheduling users with non-overlapping AOAs has already been utilized in suppressing pilot contaminant~\cite{YHF} and eliminating inter-group interference in two-stage precoding~\cite{JSDM}.
After proper user scheduling of type-S users, the interference from type-C users becomes a main factor limiting the achievable rate of type-S users.}
Besides, the interference becomes serious with the increasing number of type-C users.
Therefore, precoding schemes to eliminate the inter-user interference between type-S and type-C users are urgent.

\subsection {Mutual Interference Suppression Between Type-C and Type-S Users}
From \ref{subsection rate C} and \ref{subsection: 3.c}, it is seen that the type-C and type-S users suffer mutual interference which cannot be suppressed by the heuristic SBM method. Therefore, it is urgent to design novel \textcolor{black}{precoding methods} to suppress the mutual interference.
In this subsection, we analyze the viability of mutual interference suppression in massive MIMO systems.

\textcolor{black}{As shown in \cite{YHF}, the rank of covariance matrix $\mathbf{\Phi}$~satisfies}
\begin{equation}
\hspace{-0.2em}\mathrm{rank}( {{\mathbf{\Phi}} } ) \leq  \left( \mathrm{cos}\left( \theta ^\mathrm{min} \right) -\mathrm{cos}\left(\theta ^\mathrm{max}  \right) \right) \hspace{-0.2em} \frac{d}{\lambda} M, \; \hspace{-0.3em}\text{as} \; \hspace{-0.2em}M \to \infty,
\end{equation}
\textcolor{black}{where $\left[ \theta ^\mathrm{min},  \theta ^\mathrm{max}  \right]$ denotes the AOA interval satisfying~$\theta ^\mathrm{min} \leq  \theta ^\mathrm{max}$, the notation $\lambda$ is wavelength and $d$ is the antenna spacing at the BS.}
It is indicated that the covariance matrices of  users are rank deficient, which means that the dominant eigenmodes of the covariance matrices are limited.
Thus, massive MIMO systems have large null spaces which could be exploited to simultaneously serve several type-S users with non-overlapping AOAs and without suffering interference.
\textcolor{black}{Furthermore, type-C users could avoid causing interference to type-S users by
sacrificing some channel information in the space spanned by the dominant eigenmodes of type-S users and bearing a little performance loss.}

Therefore, massive MIMO systems have the ability to simultaneously serve type-C and type-S users \textcolor{black}{without mutual interference}. In Section~\ref{section: eZF precoding} and~\ref{section: precoding MRT}, we will propose two novel precoding methods to suppress the mutual interference.

\section{The Proposed Extended Zero-forcing Precoding Method}\label{section: eZF precoding}

We assume that the BS holds the estimated instantaneous CSI of type-C users and the statistical CSI of type-S users.
To suppress the mutual interference between type-C and type-S users, the~eZF precoding method is proposed in this section with the purpose of minimizing the total transmit power of the BS.
The problem formulation and the precoding designs for each type of users are given in the following subsections.

\subsection{Problem Formulation}
Due to the absence of type-S users' instantaneous CSI, we consider the suppression of average interference to type-S users in this subsection. Based on the ZF criterion, the optimization problem for the eZF method is formulated as
\vspace{-0.5em}
\begin{subequations}\label{eq 3.1-5}
\begin{align}
\underset{{\mathbf{W}_\mathrm{C}},{\mathbf{W}_\mathrm{S}}}{\text{minimize}}  \quad &||{\mathbf{W}_\mathrm{C}}||_F^2+||{\mathbf{W}_\mathrm{S}}||_F^2  \label{eq 3.001} \\
\text{subject\;to} \quad  &{\widehat{\mathbf{G}}^H_{\mathrm{C}}}{\mathbf{W}_{\mathrm{C}}} =\sqrt{\rho} \mathbf{I}_K, \label{eq 3.002} \\
& { \mathbf{W}^H_\mathrm{C}}{ {\mathbf{\Phi}}_{\mathrm{S},n}}{\mathbf{W}_\mathrm{C}}=\mathbf{0}_K, \forall n , \label{eq 3.003} \\
& {\mathbf{w}^H_{\mathrm{S},n}}{ {\mathbf{\Phi}}_{\mathrm{S},n}}{\mathbf{w}_{\mathrm{S},n}}= \rho, \forall n , \label{3.004} \\
&{\mathbf{w}^H_{\mathrm{S},n}}{\widehat{\mathbf{G}}_\mathrm{C}}{\widehat{\mathbf{G}}^H_\mathrm{C}}{\mathbf{w}_{\mathrm{S},n}}=0,\forall n ,\label{eq 3.005} \\
&{\mathbf{w}^H_{\mathrm{S},n}}{ {\mathbf{\Phi}}_{\mathrm{S},i}}{\mathbf{w}_{\mathrm{S},n}}=0,\; i \neq n, \forall n ,\label{eq 3.006}
\end{align}
\end{subequations}
\textcolor{black}{where $\widehat{\mathbf{G}}_\mathrm{C} \triangleq {\left[ {\begin{array}{*{20}{c}}
{{\widehat{\mathbf{g}}_{\mathrm{C},1}}}, & \cdots &, {{\widehat{\mathbf{g}}_{\mathrm{C},K}}}
\end{array}} \right]}\in \mathbb{C}^{M \times K} $ is the estimated channel matrix of type-C users}, and
the objective function (\ref{eq 3.001}) denotes the total transmit power of the BS.
The constraint~(\ref{eq 3.002}) implies that the interference among type-C users is cancelled.
Moreover, \textcolor{black}{the received signal power of each type-C and type-S users is set to $\rho$ in (\ref{eq 3.002}) and~(\ref{3.004}), respectively.
The parameter $\rho$ is to ensure the quality of service for users when the transmit power is controlled.}
The constraints~(\ref{eq 3.003}), (\ref{eq 3.005}) and (\ref{eq 3.006}) aim to eliminate the average interference from type-C to type-S users, the interference from type-S to type-C users and the average interference among type-S users, respectively.
Moreover, it is observed that $\mathbf{W}_\mathrm{C}$ and ${\mathbf{w}_{\mathrm{S},n}}, \forall n=1,\dots,N, $ in~(\ref{eq 3.1-5}) are not coupled. \textcolor{black}{Therefore, the optimal solution $\mathbf{W}^{\mathrm{eZF}}_\mathrm{C}$ and ${\mathbf{w}^{\mathrm{eZF}}_{\mathrm{S},n}} $ could be obtained by solving corresponding sub-problems of~(\ref{eq 3.1-5}), which will be given as follows.}

\subsection{Precoding Design for Type-C users}\label{subsection: ezf c}
Since $\mathbf{W}_\mathrm{C}$ and ${\mathbf{w}_{\mathrm{S},n}},\forall n=1,\dots ,N,$ in (\ref{eq 3.1-5}) are independent, the optimal precoding matrix~$\mathbf{W}^{\mathrm{eZF}}_\mathrm{C}$ for type-C users is designed by optimizing a sub-problem of (\ref{eq 3.1-5}) as
\begin{subequations}\label{eq 3.c}
\begin{align}
\underset{{\mathbf{W}_\mathrm{C}} }{\text{minimize}} \quad &||{\mathbf{W}_\mathrm{C}}||_F^2\label{eq 3.1}\\
\text{subject\;to}  \quad & {\widehat{\mathbf{G}}^H_{\mathrm{C}}}{\mathbf{W}_{\mathrm{C}}} = \sqrt{\rho} \mathbf{I}_K,\label{eq n 3.22}\\
&{\mathbf{W}^H_\mathrm{C}}{ {\mathbf{\Phi}}_{\mathrm{S},n}}{\mathbf{W}_\mathrm{C}}=\mathbf{0}_K, \forall n .\label{eq 3.3}
\end{align}
\end{subequations}
Firstly, we define ${\mathbf{\Phi}_\mathrm{S}} \triangleq \sum\limits_{n = 1}^N {\mathbf{\Phi}_{\mathrm{S},n}}$.
Since ${\mathbf{\Phi}_{\mathrm{S},n}}, \forall n$ is positive semi-definite, the value of the left-hand side of (\ref{eq 3.3}) must be a non-negative vector. Therefore, when~$\mathbf{W} _\mathrm{C}$ satisfies
\begin{equation}\label{eq n 1}
{\mathbf{W}^H_\mathrm{C}}{ {\mathbf{\Phi}}_{\mathrm{S}}}{\mathbf{W}_\mathrm{C}}=\mathbf{0}_K,
\end{equation}
the precoding matrix $\mathbf{W} _\mathrm{C}$ must satisfy the constraint (\ref{eq 3.3}).
Let~${\mathbf{\Phi}_\mathrm{S}}=\mathbf{U}_1 \mathbf{\Lambda}_1 \mathbf{U}_1^H$ be the eigen-decomposition of~the symmetric matrix~${ {\mathbf{\Phi}}_{\mathrm{S}}}$, where ${{ {\mathbf{\Lambda}}_{1}}}$ is a diagonal
matrix composed by eigenvalues in decreasing order and ${\mathbf{U}}_{1}$ is a unitary matrix.
Then, the equation (\ref{eq n 1}) becomes
\begin{eqnarray}\label{eq n 3}
{\mathbf{W}^H_\mathrm{C}}{\mathbf{U}_1 \mathbf{\Lambda}_1 \mathbf{U}_1^H}{\mathbf{W}_\mathrm{C}}={\overline {\mathbf{W}}^H _{\mathrm{C}}}   \mathbf{\Lambda}_1 {\overline {\mathbf{W}} _{\mathrm{C}}} =\mathbf{0}_K,
\end{eqnarray}
where ${\overline {\mathbf{W}} _{\mathrm{C}}}  \triangleq \mathbf{U}_1^H \mathbf{W}_\mathrm{C} $.
When the rank of ${\mathbf{\Phi}_\mathrm{S}}$ is denoted as~$r_1  \triangleq \mathrm{rank} ({\mathbf{\Phi}_\mathrm{S}})$, the first~$r_1$ eigenvalues in $\mathbf{\Lambda}_1$ are non-zero.
Then, the first $r_1$ rows of $\overline {\mathbf{W}} _{\mathrm{C}}$ must be a $r_1 \times K$ zero matrix to satisfy the condition in (\ref{eq n 3}), such as
 \begin{equation}\label{eq simu 1}
\overline {\mathbf{W}} _{\mathrm{C}}= {\left[ {\begin{array}{*{20}{c}}
{\mathbf{0}_{{r_1} \times K}^T}&{{{\overline {\mathbf{W}}^T _{\mathrm{C}1}}}}
\end{array}} \right]^T}.
 \end{equation}
Secondly, due to $ \mathbf{U}_1 \mathbf{U}_1^H=\mathbf{I}_{M} $, the constraint (\ref{eq n 3.22}) is rewritten as
\begin{eqnarray}\label{eq1}
{\widehat{\mathbf{G}}^H_{\mathrm{C}}} \mathbf{U}_1 \mathbf{U}_1^H {\mathbf{W}_{\mathrm{C}}}  ={\overline {\mathbf{G}}^H _\mathrm{C}} {\overline {\mathbf{W}} _{\mathrm{C}}} = \sqrt{\rho} \mathbf{I}_K,
\end{eqnarray}
where ${\overline {\mathbf{G}} _\mathrm{C}}  \triangleq \mathbf{U}_1^H {\widehat{\mathbf{G}}_\mathrm{C}} $. By combining (\ref{eq simu 1}) with (\ref{eq1}) and dividing the matrix ${\overline {\mathbf{G}} _\mathrm{C}}$ into two matrices as~${\overline {\mathbf{G}} _\mathrm{C}}  \triangleq \left[ {\begin{array}{*{20}{c}}
{{{\overline {\mathbf{G}}}^T _{\mathrm{C}1}}}&{{{\overline {\mathbf{G}} }^T_{\mathrm{C}2}}}
\end{array}} \right]^T$ with ${\overline {\mathbf{G}} }_{\mathrm{C}1} \in \mathbb{C}^{r_1 \times K}$ and ${\overline {\mathbf{G}} }_{\mathrm{C}2} \in \mathbb{C}^{{\left( M-r_1 \right)} \times K}$,
the constraint (\ref{eq n 3.22}) equals~to
\begin{eqnarray}\label{eq11}
{\overline {\mathbf{G}} }^H_{\mathrm{C}2} {\overline {\mathbf{W}} _{\mathrm{C}1}} =\sqrt{\rho} \mathbf{I}_K.
\end{eqnarray}
Furthermore, based on (\ref{eq simu 1}), the objective function in (\ref{eq 3.1}) is rewritten as
\begin{equation}
\left\|  {\mathbf{W}_\mathrm{C}} \right\|^2_F=\left\| \overline {\mathbf{W}} _{\mathrm{C1}} \right\|^2_F.
\end{equation}
Hence, the optimization problem (\ref{eq 3.c}) is equivalent to
\begin{subequations}\label{eq 4.c}
\begin{align}
\underset{\overline {\mathbf{W}} _{\mathrm{C1}} }{\text{minimize}} \quad &||\overline {\mathbf{W}} _{\mathrm{C1}}||_F^2  \\
 \text{subject\; to}   \quad  &{\overline {\mathbf{G}} }^H_{\mathrm{C}2} {\overline {\mathbf{W}} _{\mathrm{C}1}}= \sqrt{\rho} \mathbf{I}_K.\label{eq 3.22}
\end{align}
\end{subequations}
 Since the optimization problem (\ref{eq 4.c}) has a similar form to the
ZF criterion, based on the properties of
generalized inverses in linear algebra \cite{Ami2008}, the optimal solution
to~(\ref{eq 4.c})~is
\begin{equation}
\overline {\mathbf{W}}^{\mathrm{opt}} _{\mathrm{C1}}=\sqrt{\rho} {{\overline {\mathbf{G}} }_{\mathrm{C}2}^\dag }.
\end{equation}
Then, the optimal solution to (\ref{eq 3.c}) is
\begin{equation}\label{eq si 1}
\mathbf{W}_\mathrm{C}^\mathrm{eZF}=\mathbf{U}_1 {\overline {\mathbf{W}}^{\mathrm{opt}} _{\mathrm{C}}} = \sqrt{\rho} \mathbf{U}_1  \left[ {\begin{array}{*{20}{c}}
{{\mathbf{0}_{{r_1} \times K}}}\\
{  { {{\overline {\mathbf{G}} }_{\mathrm{C}2}} ^\dag } }
\end{array}} \right].
\end{equation}

\textcolor{black}{
It is seen from (\ref{eq n 1}) that the type-S users mainly exist in a subspace of $r_1$ dimensions.
Due to the large scale of antenna elements, massive MIMO systems have much larger degrees of freedom (DoF) than $r_1$.
Intuitively, the excess DoF could be used to suppress the interference among users \cite{Tat2015}.
Then, the type-C users could avoid causing interference to the type-S users at the cost of reducing array gain to $M-r_1$.
Thus, the type-C users will suffer performance loss with respect to the conventional transmission mode only serving type-C users.
Nevertheless,
when the number of BS antennas is large enough, $r_1$ becomes negligible compared to $M$ and the performance loss of type-C users could be attenuated. This will be verified by the simulation results in Section \ref{section:simu}.}

\subsection{Precoding Design for type-S users}\label{subsection: ezf s}
As a sub-problem of (\ref{eq 3.1-5}), the optimal precoding vector~$\mathbf{w}^{\mathrm{eZF}}_{\mathrm{S},n}$ for the $n$-th type-S user is obtained~by
\vspace{-0.5em}
\begin{subequations}\label{eq 444.0}
\begin{align}
\underset{ {\mathbf{w}_{\mathrm{S},n}} }{\text{minimize}} \quad &||{\mathbf{w}_{\mathrm{S},n}}||^2\label{eq 4.1}\\
\text{subject\;to}  \quad&{\mathbf{w}^H_{\mathrm{S},n}}{ {\mathbf{\Phi}}_{\mathrm{S},n}}{\mathbf{w}_{\mathrm{S},n}}= {\rho},\forall n ,\label{24.004}\\
&{\mathbf{w}^H_{\mathrm{S},n}}{\widehat{\mathbf{G}}_\mathrm{C}} {\widehat{\mathbf{G}}^H_\mathrm{C}}{\mathbf{w}_{\mathrm{S},n}}=0,\forall n ,\label{eq 224.005}\\
&{\mathbf{w}^H_{\mathrm{S},n}}{ {\mathbf{\Phi}}_{\mathrm{S},i}}{\mathbf{w}_{\mathrm{S},n}}=0,\; i \neq n,\forall n .\label{eq 224.006}
\end{align}
\end{subequations}
Since ${\widehat{\mathbf{G}}^H_\mathrm{C}}{\widehat{\mathbf{G}}_\mathrm{C}}$ and $ {\mathbf{\Phi}}_{\mathrm{S},i}, i \neq n,$ are both positive semi-definite, the constraints (\ref{eq 224.005}) and (\ref{eq 224.006}) equal to
\begin{equation} \label{heuheu24}
\mathbf{w}^H_{\mathrm{S},n}  \big ( {{\widehat{\mathbf{G}}_\mathrm{C}}{\widehat{\mathbf{G}}^H_\mathrm{C}}+ { {\sum\limits _{i=1,i \neq n}^N \hspace{-2mm} { {\mathbf{\Phi}}_{\mathrm{S},i}} }}} \big ){\mathbf{w}_{\mathrm{S},n}}=0.
\end{equation}
Due to the similarity between (\ref{heuheu24}) and (\ref{eq 3.3}), we define~$\mathbf{Q}_n \triangleq {{\widehat{\mathbf{G}}_\mathrm{C}}{\widehat{\mathbf{G}}^H_\mathrm{C}}+  { {\sum\limits _{i=1,i \neq n}^N { {\mathbf{\Phi}}_{\mathrm{S},i}} }}}$.
The rank of $\mathbf{Q}_n$ is denoted as~$r_2$ and its eigen-decomposition is given as~${\mathbf{Q}_n}=\mathbf{U}_2 {\mathbf{\Lambda}_2} \mathbf{U}_2^H$.
Then, following the procedures from (\ref{eq n 1}) to (\ref{eq simu 1}), we have
 \begin{equation}\label{1.10}
{\overline {\mathbf{w}} _{\mathrm{S},n}}=\textcolor{black}{\mathbf{U}^H_2 {\mathbf{w}} _{\mathrm{S},n}} = {\left[ {\begin{array}{*{20}{c}}
{\mathbf{0}_{{r_2} \times 1}^T}&{{{\overline {\mathbf{w}}^T _{\mathrm{S1},n}}}}
\end{array}} \right]^T}.
 \end{equation}
Set ${\overline {\mathbf{\Phi}} _{\mathrm{S},n}}\triangleq {\mathbf{U}_2^H} {\mathbf{\Phi}} _{\mathrm{S},n} \mathbf{U}_2$ and rewrite it as $\textcolor{black}{{\overline {\mathbf{\Phi}} _{\mathrm{S},n}} \triangleq \left[ {\begin{array}{*{20}{c}}
{\overline {\mathbf{\Phi}} _{\mathrm{S1},n }}&{\mathbf{{C}}_1}\\
{{\mathbf{{C}}_2}}&{{\overline {\mathbf{\Phi}} _{\mathrm{S2},n }}}
\end{array}} \right]}$ with $ \overline {\mathbf{\Phi}} _{\mathrm{S1},n} \in \mathbb{C}^{r_2 \times r_2}$ and~$ {\overline {\mathbf{\Phi}} _{\mathrm{S2},n}} \in \mathbb{C}^{(M-r_2) \times (M-r_2)}$.
The constraint (\ref{24.004}) is equivalent to
\begin{equation}\label{eq 20170521.2}
\overline {\mathbf{w}}^H _{\mathrm{S1},n}  \overline {\mathbf{\Phi}} _{\mathrm{S2},n}  \overline {\mathbf{w}} _{\mathrm{S1},n}=  {\rho}.
\end{equation}
Since the objective function in (\ref{eq 4.1}) is equivalent to
$\left\|{\mathbf{w}_{\mathrm{S},n}}\right\|^2=\left\|  \overline {\mathbf{w}} _{\mathrm{S1},n } \right\|^2$,
the optimization problem (\ref{eq 444.0}) is transformed to
\begin{subequations}\label{eq heu s 3}
\begin{align}
\underset{ \overline {\mathbf{w}} _{\mathrm{S1},n }  }{\text{minimize}} \quad & \left\|  \overline {\mathbf{w}} _{\mathrm{S1},n } \right\|^2 \\
\text{subject\;to} \quad & \overline {\mathbf{w}}^H _{\mathrm{S1},n }  \overline {\mathbf{\Phi}} _{\mathrm{S2},n}   \overline {\mathbf{w}} _{\mathrm{S1},n }= \rho,\forall n.\label{eq heu s 2}
\end{align}
\end{subequations}
The eigen-decomposition of $  \overline {\mathbf{\Phi}} _{\mathrm{S2},n }$ is $ \overline {\mathbf{\Phi}} _{\mathrm{S2},n }= \overline {\mathbf{U}}_\mathrm{S} { \overline {\mathbf{\Lambda}}}_\mathrm{S} \overline {\mathbf{U}}^H_\mathrm{S}$.
Let vector~$ {\mathbf{u}}_{\mathrm{S}i} ,\forall i=1,\cdots,M ,$ denote the~$i$-th column of $ \overline {\mathbf{U}}_\mathrm{S} $ and $\lambda _{\mathrm{S}i},\forall i=1,\cdots,M,$ denote the corresponding eigenvalue. Then, since the design of $\overline {\mathbf{w}} _{\mathrm{S1},n}$ is based on~$ \overline {\mathbf{\Phi}} _{\mathrm{S2},n}  $, the vector~$\overline {\mathbf{w}} _{\mathrm{S1},n}$
 is a linear combination of the column vectors of $\overline {\mathbf{U}}_\mathrm{S}$, i.e.,
$\overline {\mathbf{w}} _{\mathrm{S1},n} = {\omega _{1}}{\overline{{\mathbf{u}}}_{\mathrm{S}1}}+\omega _{2}{{\overline{\mathbf{u}}}_{\mathrm{S}2}}+\cdots   +{\omega _{M}}{{\overline{\mathbf{u}}}_{\mathrm{S}M}}$.
Thus, the optimization problem (\ref{eq heu s 3}) becomes
\begin{subequations}\label{eq heu s 4}
\begin{align}
\underset{ \omega_{i} }{\text{minimize}}  \quad &\sum\limits_{i = 1}^M {|{\omega_{i}}{|^2}}\label{eq 4.10}\\
\text{subject\;to}  \quad& \sum\limits_{i = 1}^M {|{\omega_{i}}{|^2}{\lambda _{\mathrm{S}i}}}  =  {\rho},\forall i .\label{eq 4.11}
\end{align}
\end{subequations}
When the eigenvalues $ \lambda _{\mathrm{S}i}, \forall i=1,\cdots,M,  $ are sorted in decreasing order, the optimal solution to the problem (\ref{eq heu s 4}) is~${\omega _{ 1}} = \sqrt {\frac{ {\rho}}{{{\lambda _{\mathrm{S}1}}}}},\;{\omega _{ 2}} =  \cdots  = {\omega _{ M}} = 0 $.
Thus, the optimal~$\overline {\mathbf{w}} _{\mathrm{S1},n }$~is
 \begin{align}\label{eq 4.8}
 \overline {\mathbf{w}}^{\mathrm{opt}} _{\mathrm{S1},n }& ={\sqrt {\frac{{\rho}}{{{\lambda _{\mathrm{S}1}}}}} }{\mathbf{u}_{\mathrm{S}1} } ={\sqrt {\frac{{\rho}}{{{\lambda _{ \mathrm{max}}}\big( \overline {\mathbf{\Phi}} _{\mathrm{S2},n } \big)}}} }{\mathbf{u}_\mathrm{max}\big( \overline {\mathbf{\Phi}} _{\mathrm{S2},n } \big)}.
\end{align}
Finally, the proposed eZF precoding vector for the $n$-th type-S user is designed as
\begin{align} \label{t2}
\mathbf{w}_{\mathrm{S},n}^\mathrm{eZF} &=\mathbf{U}_2 {\overline {\mathbf{w}}^{\mathrm{opt}} _{\mathrm{S},n}} \nonumber \\
&= \sqrt {\frac{{\rho}}{{{\lambda _{ \mathrm{max}}}\big( \overline {\mathbf{\Phi}} _{\mathrm{S2},n } \big)}}}  \mathbf{U}_2  \left[ {\begin{array}{*{20}{c}}
{{\mathbf{0}_{{r_2} \times 1}}} \\
{\mathbf{u}_\mathrm{max}\big( \overline {\mathbf{\Phi}} _{\mathrm{S2},n } \big)}
\end{array}} \right].
\end{align}

As shown in (\ref{eq 20170521.2}), the $n$-th type-S user has to sacrifice part of useful channel statistical information to avoid causing interference to other users, which leads to a degradation of type-S users' average received signal power.

\section{The Proposed Extended Maximum Ratio Transmission Precoding Method}\label{section: precoding MRT}

\textcolor{black}{
Different from the eZF method, the eMRT method is proposed in this section, which aims at maximizing the received signal power of each user as well as eliminating the mutual interference between the type-C and type-S users.
The problem formulation of the eMRT method and precoding matrix design are presented in the following subsections.}

\subsection{Problem Formulation}
By considering the mutual interference suppression between  type-S and type-C users, the proposed eMRT precoding method is formulated as
\vspace{-0.5em}
\begin{subequations}\label{eq 5.40}
\begin{align}
\!\!\!\!\!\!\!\!\!\!\!\!  \underset{\mathbf{W}_\mathrm{C}, \mathbf{w}_{\mathrm{S},n}}{\text{maximize}}& \quad  ||\widehat{\mathbf{G}}^H_\mathrm{C}\mathbf{W}_\mathrm{C}||_F^2 +  \sum\limits_{n=1}^N  {\mathbf{w}_{\mathrm{S},n}^H}{ {\mathbf{\Phi}}_{\mathrm{S},n}}{\mathbf{w}_{\mathrm{S},n}}\label{eq 5.414}\\
\text{subject\;to} & \quad||\mathbf{w}_{\mathrm{C},k}||^2=p_d,\forall k ,\label{eq 5.5}\\
&\quad||\mathbf{w}_{\mathrm{S},n}||^2= p_d,\forall n ,\label{eq 5.7}\\
&\quad\mathbf{W}_\mathrm{C}^H{\mathbf{\Phi}_{\mathrm{S},n}}{\mathbf{W}_\mathrm{C}}=\mathbf{0}_K,\forall n ,\label{eq 5.6}\\
&\quad{\mathbf{w}^H_{\mathrm{S},n}}{\widehat{\mathbf{G}}^H_\mathrm{C}}{\widehat{\mathbf{G}}_\mathrm{C}}{\mathbf{w}_{\mathrm{S},n}}=0, \forall n.\label{eq 5.88}
\end{align}
\end{subequations}
The average received signal power of type-S users is considered in the objective function~(\ref{eq 5.414}) since the instantaneous CSI of type-S users is unknown to the BS.
The constraints~(\ref{eq 5.5}) and (\ref{eq 5.7}) indicate that the transmit power from the BS to each type-C and type-S users are set to $p_d$.
The constraint~(\ref{eq 5.6}) is to eliminate the average interference power from type-C users to type-S users.
The constraint~(\ref{eq 5.88}) is to eliminate the interference power from type-S users to type-C users.
Since~$\mathbf{w}_{\mathrm{S},n}, \forall n=1, \dots ,N,$ and $\mathbf{W}_\mathrm{C}$ in (\ref{eq 5.40}) are not coupled, they could be solved separately in the following subsections.

\subsection{Precoding Design for Type-C users}
The sub-problem of (\ref{eq 5.40}) to obtain the optimal precoding matrix $\mathbf{W}^{\mathrm{eMRT}}_\mathrm{C}$ for type-C users is
\begin{subequations}\label{eq 3322}
\begin{align}
\underset{\mathbf{W}_\mathrm{C}  }{\text{maximize}} \quad &||\widehat{\mathbf{G}}^H_\mathrm{C}\mathbf{W}_\mathrm{C}||_F^2\label{eq 555.4}\\
\text{subject\;to}  \quad & ||\mathbf{w}_{\mathrm{C},k}||^2=p_d,\forall k ,\label{b1} \\
&\mathbf{W}_\mathrm{C}^H{\mathbf{\Phi}_{\mathrm{S},n}}{\mathbf{W}_\mathrm{C}}=\mathbf{0}_K,\forall n .\label{eq e555.7}
\end{align}
\end{subequations}
Following a similar \textcolor{black}{derivation} of the constraint (\ref{eq 3.3}), the problem~(\ref{eq 3322}) is converted to
\begin{subequations}\label{eq heu p1}
\begin{align}
\underset{\overline {\mathbf{W}} _{\mathrm{C}1} }{\text{maximize}}  \quad &||{\overline {\mathbf{G}} }^H_{\mathrm{C}2} \overline {\mathbf{W}} _{\mathrm{C}1}||_F^2\label{eq 5.4}\\
\text{subject\;to} \quad & \left\| {\overline {\mathbf{w}} _{\mathrm{C1},k}} \right\|^2=p_d ,\forall k,
\end{align}
\end{subequations}
where ${\overline {\mathbf{w}} _{\mathrm{C1},k}} $ is the $k$-th column of $\overline {\mathbf{W}} _{\mathrm{C}1}$. Since the optimization problem (\ref{eq heu p1}) has \textcolor{black}{a} similar form to the MRT criterion, its optimal solution is
\begin{equation}
{\overline {\mathbf{w}}^{\mathrm{opt}} _{\mathrm{C1},k}}=\sqrt{p_d} \frac{{\overline {\mathbf{g}} _{\mathrm{C2},k}}}{||  {\overline {\mathbf{g}} _{\mathrm{C2},k}} ||},
\end{equation}
where $ {\overline {\mathbf{g}} _{\mathrm{C2},k}} $ is the $ k $-th column of $ {\overline {\mathbf{G}} _{\mathrm{C2}}} $ given after (\ref{eq1}).
Due to~$\overline {\mathbf{W}} _{\mathrm{C}}={\mathbf{U}}^H _1 {\mathbf{W}} _{\mathrm{C}} = {\left[ {\begin{array}{*{20}{c}}
{\mathbf{0}_{{r_1} \times K}^T}&{{{\overline {\mathbf{W}}^T _{\mathrm{C1}}}}}
\end{array}} \right]^T}$, the optimal eMRT precoding vector for the $ k $-th type-C user is given as\footnote{
\textcolor{black}{The matrix $\mathbf{\Phi}_{\mathrm{S},n}$ in (\ref{eq 3.c}) and (\ref{eq 3322}) might be full rank in practice.
Then, we have~$r_1=M$ and cannot obtain a non-zero solution for $\mathbf{W}_\mathrm{C}^\mathrm{eZF}$/$\mathbf{W}_\mathrm{C}^\mathrm{eMRT}$. To overcome this problem,
low rank approximation for matrices could be used to obtain an optimal rank-$D_n$ approximation of~$\mathbf{\Phi}_{\mathrm{S},n}$~\cite{low-rank}.
Thus, the interference from type-C users to type-S users existing in the subspace of $D_n$-dimension is removed. Note that $D_n$ is adjustable to realistic scenarios and requirements, which could be determined by measurements and~tests.}
}
\begin{equation}\label{eq si 2}
{\mathbf{w}}^\mathrm{eMRT} _{\mathrm{C},k}=  \mathbf{U} _1 {\overline {\mathbf{w}} _{\mathrm{C1},k}}=\frac{\sqrt{p_d} }{ ||  {\overline {\mathbf{g}} _{\mathrm{C2},k}} || } \mathbf{U} _1 \left[ {\begin{array}{*{20}{c}}
{{\mathbf{0}_{{r_1} \times 1}}}\\
{ {\overline {\mathbf{g}}  _{\mathrm{C2},k}} }
\end{array}} \right].
\end{equation}
\textcolor{black}{
Note that the type-C users sacrifice some channel information and suffer performance loss to avoid causing interference to type-S users. Fortunately, we will find that the performance loss of type-C users decreases with the increasing number of BS antennas in Section \ref{section:simu}.}

\begin{table*}[ht]
\caption{Summary of the precoding vectors/matrcies for type-C and type-S users} \label{summary}
\centering 
\begin{tabular}{|c|c|c|}
\hline
\hline
{} & Type-C users & Type-S users \\
\hline
\hline
{SBM} & $\textbf{w}_{\mathrm{C},k}^\mathrm{SBM}=\sqrt{p_d}  \frac{\widehat{\mathbf{g}}_{{\mathrm{C},k}} }{\left\| {\widehat{\mathbf{g}}_{{\mathrm{C},k}}} \right\|}$ (\ref{heu c 1}) & $\textbf{w}_{\mathrm{S},n}^\mathrm{SBM}=\sqrt{p_d } \mathbf{u}_\mathrm{max} \left ( \mathbf{\Phi}_{\mathrm{S},n} \right)$ (\ref{heuheu9})\\ \hline
{eZF } & $\mathbf{W}_\mathrm{C}^\mathrm{eZF}  = \hspace{-1.2mm} \sqrt{\rho} \mathbf{U}_1  \left[ {\begin{array}{*{20}{c}}
{{\mathbf{0}_{{r_1} \times K}}}\\
{  { {{\overline {\mathbf{G}} }_{\mathrm{C}2}} ^\dag } }
\end{array}} \right]$  (\ref{eq si 1}) &$\mathbf{w}_{\mathrm{S},n}^\mathrm{eZF}  = \hspace{-1.8mm} \sqrt {\frac{{\rho}}{{{\lambda _{ \mathrm{max}}}\big( \overline {\mathbf{\Phi}} _{\mathrm{S2},n } \big)}}}  \mathbf{U}_2  \left[ {\begin{array}{*{20}{c}}
{{\mathbf{0}_{{r_2} \times 1}}} \\
{\mathbf{u}_\mathrm{max}\big( \overline {\mathbf{\Phi}} _{\mathrm{S2},n } \big)}
\end{array}} \right]$  (\ref{t2}) \\
\hline
{eMRT } & ${\mathbf{w}}^\mathrm{eMRT} _{\mathrm{C},k}= \hspace{-1mm} \frac{\sqrt{p_d} }{ ||  {\overline {\mathbf{g}} _{\mathrm{C2},k}} || } \mathbf{U} _1 \left[ {\begin{array}{*{20}{c}}
{{\mathbf{0}_{{r_1} \times 1}}}\\
{ {\overline {\mathbf{g}} _{\mathrm{C2},k}} }
\end{array}} \right]$  (\ref{eq si 2}) & $\mathbf{w}_{\mathrm{S},n}^\mathrm{eMRT}= \hspace{-1mm} \sqrt{p_d}\mathbf{U}_3  \left[ {\begin{array}{*{20}{c}}
{{\mathbf{0}_{{K} \times 1}}}\\
{ {\mathbf{u}_{\max }}\left( \overline{\overline {\mathbf{\Phi}}} _{\mathrm{S2},n }\right)}
\end{array}} \right]$ (\ref{t1} ) \\
\hline
\end{tabular}
\end{table*}

\subsection{Precoding Design for Type-S Users} \label{subsection: emrt s}
Based on problem (\ref{eq 5.40}), the optimal precoding vector for the $n$-th type-S user is obtained by
\begin{subequations} \label{ppp 2}
\begin{align}
\underset{\mathbf{w}_{\mathrm{S},n}}{\text{maximize}} \quad &{\mathbf{w}_{\mathrm{S},n}^H}{ {\mathbf{\Phi}}_{\mathrm{S},n}}{\mathbf{w}_{\mathrm{S},n}}\label{eq heu 12}\\
\text{subject\;to} \quad &||\mathbf{w}_{\mathrm{S},n}||^2= p_d,\forall n ,\label{eq heu 11}\\
&\mathbf{w}_{\mathrm{S},n}^H {\widehat{\mathbf{G}}_\mathrm{C}} {\widehat{\mathbf{G}}^H_\mathrm{C}}\mathbf{w}_{\mathrm{S},n}=0,\forall n .\label{eq heu s 9}
\end{align}
\end{subequations}
Since the constraint (\ref{eq heu s 9}) has a similar form to~(\ref{heuheu24}), we could~obtain
\begin{equation}\label{b2}
\textcolor{black}{{\overline{\overline {\mathbf{w}}} _{\mathrm{S},n}}=  {\mathbf{U}^H_3 } {  {\mathbf{w}} _{\mathrm{S},n}} =
{\left[ {\begin{array}{*{20}{c}}
{\mathbf{0}_{{K} \times 1}^T}&{{{\overline {\mathbf{w}}^T _{\mathrm{S2},n}}}}
\end{array}} \right]^T}}
\end{equation}
by setting $ \mathbf{P} \triangleq {\widehat{\mathbf{G}}_\mathrm{C}^H}{\widehat{\mathbf{G}}_\mathrm{C}} $, where $K$ is the rank of $ \mathbf{P}$. The matrix~$\mathbf{U}_3 $ is obtained by the eigen-decomposition of $ \mathbf{P}$, i.e.,~$ \mathbf{P}  = \mathbf{U}_3 {\mathbf{\Lambda}_3} \mathbf{U}_3^H$.

Let ${\overline{\overline {\mathbf{\Phi}}} _{\mathrm{S},n}} \triangleq \mathbf{U}_3 { {\mathbf{\Phi}}_{\mathrm{S},n}}\mathbf{U}_3^H$ and rewrite it as ${\overline{\overline {\mathbf{\Phi}}} _{\mathrm{S},n}} \triangleq {\left[ {\begin{array}{*{20}{c}}
{\overline{\overline {\mathbf{\Phi}}} _{\mathrm{S1},n }}&{\mathbf{E}_1}\\
{{\mathbf{E}_2}}&{{\overline{\overline {\mathbf{\Phi}}} _{\mathrm{S2},n }}}
\end{array}} \right]}$ with $ \overline{\overline {\mathbf{\Phi}}} _{\mathrm{S1},n } \in \mathbb{C}^{K\times  K}$ and~$ \overline{\overline {\mathbf{\Phi}}} _{\mathrm{S2},n } \in \mathbb{C}^{(M-K ) \times (M-K)}$. Then, the objective function (\ref{eq heu 12}) becomes
\begin{align}
{\mathbf{w}_{\mathrm{S},n}^H}{ {\mathbf{\Phi}}_{\mathrm{S},n}}{\mathbf{w}_{\mathrm{S},n}}
=\overline {\mathbf{w}}^H _{\mathrm{S2},n }  \overline{\overline {\mathbf{\Phi}}} _{\mathrm{S2},n } \overline {\mathbf{w}} _{\mathrm{S2},n }.
\end{align}
Due to $||\overline {\mathbf{w}} _{\mathrm{S2},n }||^2=  ||\mathbf{w}_{\mathrm{S},n}||^2$, an equivalent problem of~(\ref{ppp 2})~is
\vspace{-0.5em}
\begin{subequations}\label{heu ppp 2}
\begin{align}
\underset{\overline {\mathbf{w}} _{\mathrm{S2},n }}{\text{maximize}}  \quad & \overline {\mathbf{w}}^H _{\mathrm{S2},n }  \overline{\overline {\mathbf{\Phi}}} _{\mathrm{S2},n } \overline {\mathbf{w}} _{\mathrm{S2},n }\\
\text{subject\;to}  \quad & \left\|\overline {\mathbf{w}} _{\mathrm{S2},n }\right\|=p_d,\forall n .
\end{align}
\end{subequations}
Then, the optimal solution to (\ref{heu ppp 2}) is
$ \overline {\mathbf{w}}^{\mathrm{opt}} _{\mathrm{S2},n }=\sqrt{p_d} \mathbf{u}_\mathrm{max} \left(  \overline{\overline {\mathbf{\Phi}}} _{\mathrm{S2},n } \right)$. The optimal eMRT precoding vector for the $n$-th type-S user is
\vspace{-0.5em}
\begin{equation}\label{t1}
\mathbf{w}_{\mathrm{S},n}^\mathrm{eMRT}=\mathbf{U}_3 \textcolor{black}{{\overline{\overline {\mathbf{w}}} _{\mathrm{S},n}}} =\sqrt{p_d} \mathbf{U}_3  \left[ {\begin{array}{*{20}{c}}
{{\mathbf{0}_{{K} \times 1}}}\\
{ {\mathbf{u}_{\max }}\left( \overline{\overline {\mathbf{\Phi}}} _{\mathrm{S2},n }\right)}
\end{array}} \right].
\end{equation}

\textcolor{black}{In summary, Table \ref{summary} presents the closed-form expressions of different precoding vectors/matrices for type-C and type-S users.
The performances of these methods will be shown in the simulations.}

\section{Simulation Results }\label{section:simu}
In this section, the performances of the proposed SBM, eZF and eMRT methods are evaluated.
We consider a massive MIMO system with single cell, in which the BS is equipped with~$M$ antennas serving $K$ type-C users and~$N$ type-S users simultaneously.
\textcolor{black}{Moreover, we consider a physical channel model with a fixed number of dimensions~$L$ as in~\cite{Hoydis,YHF, Ngo_model}.
For a uniform linear array, the covariance matrices $\mathbf{\Phi}$ are given as 
$\mathbf{\Phi}=\mathbf{A} \mathbf{A}^H$ where $\mathbf{A} \triangleq {\left[ {\begin{array}{*{20}{c}}
{{\mathbf{a}({\theta }_1 )}}, & \cdots &, {{\mathbf{a}({\theta }_L)}}
\end{array}} \right]} \hspace{-1mm} \in \mathbb{C}^{M \times L}$ is composed of steering vectors given as
\begin{equation}\label{eq 1111}
\hspace{-0.15em}\mathbf{a}({\theta }_l ) \triangleq \hspace{-0.25em} \frac{1}{\sqrt{L}}\hspace{-0.35em} \left[\hspace{-0.45em} {\begin{array}{*{20}{c}}
1,&\hspace{-0.75em}{{e^{ - j2\pi \frac{d}{\lambda }\hspace{-0.25em} \cos ({\theta }_l )}}} ,& \hspace{-0.85em} \cdots ,& \hspace{-0.85em} {{e^{ - j2\pi \frac{{(M - 1)d}}{\lambda }\hspace{-0.25em} \cos ({\theta }_l )}}}
\end{array}} \hspace{-0.75em}\right]^T,
\end{equation}
where antenna spacing $d$ at the BS is assumed to satisfy $d=\frac{\lambda}{2}$ ($\lambda$ denotes wavelength), the parameter ${\theta }_l$ is the random AOA corresponding to the~$l$-th direction, which is assumed to be uniformly distributed over $\left[ \overline{\theta} -\theta_{\Delta}/2, \overline{\theta} +\theta_{\Delta}/2\right]$ where~$\overline{\theta} \in \left[ {0,\pi } \right]$ is the mean AOA of the $L$ directions and $\theta_{\Delta}$ is the angle spread.
We set $L=20$ and $\theta_\Delta=10^\circ$ for all the simulations.}

Furthermore, the instantaneous CSI of type-C users is obtained by uplink mutually orthogonal pilot sequences and we assume the pilot symbols and data are modulated with Orthogonal Frequency Division Multiplexing (OFDM) whose parameters are identical to Long Term Evolution standard: a symbol interval of~$T_s=71.4$ ms, a subcarrier spacing of~$\Delta_f = 15$ kHz, a useful symbol duration of~$T_u=1/\Delta_f  = 66.7 $ ms and a guard interval of~$T_g=T_s-T_u = 4.7$~ms.
\textcolor{black}{When the channel coherence time is chosen to~$T_c=500$~$\mu$s, there are~$T_c/T_s \approx7$ OFDM symbols in a coherence interval. The channel response is constant over $T_u/T_g=14$ consecutive sub-carriers.
Thus, when $T_{\mathrm{pilot}}$ OFDM symbols are utilized for uplink pilot transmission, the maximal length of  pilot sequences is $\tau = 14 T_{\mathrm{pilot}}$ and the BS could learn the instantaneous channels for $14 T_{\mathrm{pilot}}$ users at most~\cite{Marzetta, Ngo}. }


\begin{figure}[!h]
\subfigure[\textcolor{black}{Performances comparison of the closed-form and Monte Carlo results with the SBM method.}]{
\begin{minipage}[b]{0.47\textwidth}
\includegraphics[width=3.46in]{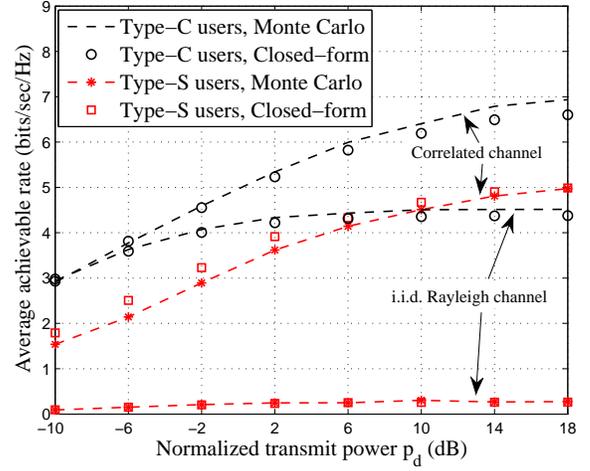}
\label{fig2_preamble}
\end{minipage}
}
\subfigure[Performances comparison of the MRT, ZF, eMRT and eZF methods.]{
\begin{minipage}[b]{0.47\textwidth}
\includegraphics[width=3.5in]{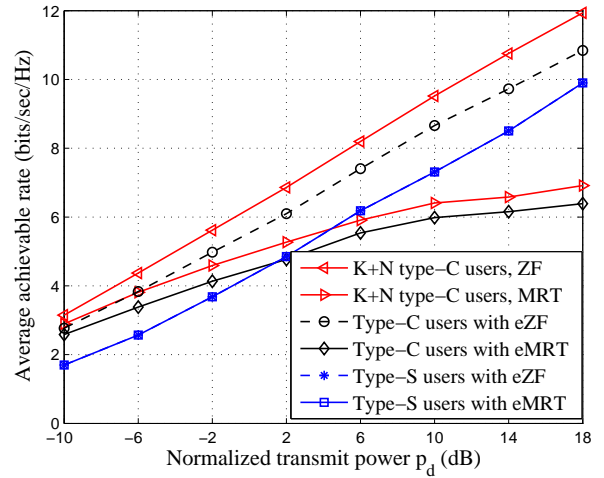}
\label{fig3 preamble}
\end{minipage}
}
 \caption{Average achievable rate for type-C and type-S users with~$M=100$, $ K=5$, $ N=1$, \textcolor{black}{$p_u$=10 dB, $T_{\mathrm{pilot}}=1$ OFDM symbol for uplink pilot transmission and MMSE receiver for type-C users.}}
\label{fig23_preamble}
\end{figure}

\textcolor{black}{We first take the downlink average achievable rate of the type-C and type-S users as a metric to illustrate the influence of introducing type-S users, which are given as~${\overline R} _\mathrm{C}^{\mathrm{prop}} \buildrel \Delta \over = \frac{1}{K} {\sum\limits_{k = 1}^K {{R_{\mathrm{C},k}}} }\; \textrm{and} \quad {\overline R }_\mathrm{S}^{\mathrm{prop}} \buildrel \Delta \over = \frac{1}{N}{\sum\limits_{n = 1}^N {{R_{\mathrm{S},n}}} },$~respectively.}
Firstly, Fig. \ref{fig2_preamble} depicts the average achievable rate with the heuristic SBM method.
It is shown that the approximate closed-form expressions derived in Section III.B and Section III.C are tight when compared to the Monte Carlo results.
Fig. \ref{fig2_preamble} also verifies that the expression given in \emph{Lemma}~1 holds for finite numbers of BS~antennas.
Moreover, it is seen from Fig.~\ref{fig2_preamble} that the type-S users with i.i.d. Rayleigh channel achieve extremely poor performance.
Hence, the BS prefers to schedule the type-S users with strong channel correlation in~practice.

To evaluate the capability of the proposed eZF and eMRT methods on mutual interference suppression, the average achievable rates for each type of users are also shown in Fig.~\ref{fig3 preamble}.
Compared to the performance of the SBM method, it is obvious that the achievable rates of users are significantly improved with the eZF and eMRT methods.
Moreover, the performances of the systems with the  conventional MRT and ZF methods serving $K+N$ users are also depicted in red lines. \textcolor{black}{In this case, the BS knows all the estimated instantaneous CSI of users and the average achievable rate of users is given as ${\overline R}^{\mathrm{conv}} \buildrel \Delta \over = \frac{1}{K+N} {\sum\limits_{i = 1}^{K+N} {{R_i}} }$.}
Since the type-C users have to sacrifice some instantaneous CSI to avoid causing interference to the type-S user, the average rate of the type-C users with the proposed precoding methods decreases slightly compared to the conventional methods.
Note that the eZF and eMRT methods achieve identical performances for the type-S user when only one type-S user is considered.


\begin{figure}[!h]
\centering
\includegraphics[width=3.5in]{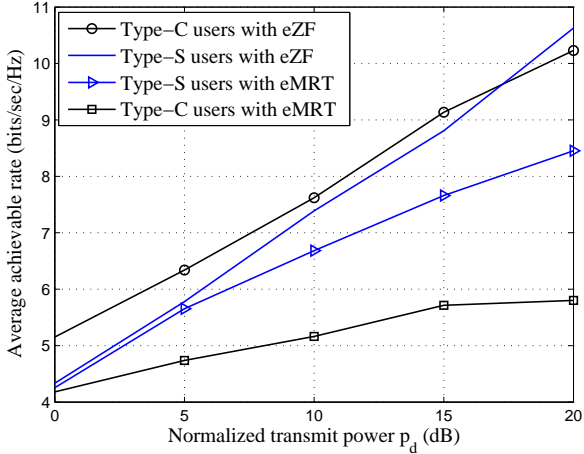}
\caption{ Average achievable rate for type-C and type-S users with $M=200$,  $ K=5$, $ N=5$, $p_u$=10 dB, $T_{\mathrm{pilot}}=1$ OFDM symbol for uplink pilot transmission and MMSE receiver for type-C users.}
\label{fig4_preamble}
\end{figure}

Fig. \ref{fig4_preamble} shows the performances of the eZF and eMRT methods with $K=5$ and $N=5$.
In this simulation, \textcolor{black}{the mean AOAs of the~$N$ type-S users are assumed to be~$ \varsigma +\frac{2 \pi}{N} n, n=0,\dots,N-1$, where $\varsigma \in [0, \pi]$ denotes the mean AOA of the first type-S user.
Thus, the type-S users have non-overlapping AOAs when the angle spread~$\theta_{\Delta}$ is~$10^\circ$. Besides, the type-C users are randomly distributed in the cell.}
It is seen from Fig. \ref{fig4_preamble} that the eZF method outperforms the eMRT method.
\textcolor{black}{From transmit power perspective, the type-C users with the eMRT method achieve the average rate of 5.5 bits/sec/Hz at 13.0 dB, while the eZF method only needs 2.0 dB.
This implies that the eZF method saves transmit power which is consistent with the purpose of the eZF method, i.e., minimizing the transmit~power.
Moreover, when the eZF method is adopted, the type-S users are able to obtain comparable average achievable rate with that of type-C users.
When the eMRT method is adopted, the type-S users even achieve better performance than type-C users.
Therefore, precoding for type-S users with only statistical CSI is a promising scheme in massive MIMO systems.}

\begin{figure}[!h]
\centering
\includegraphics[width=3.5in]{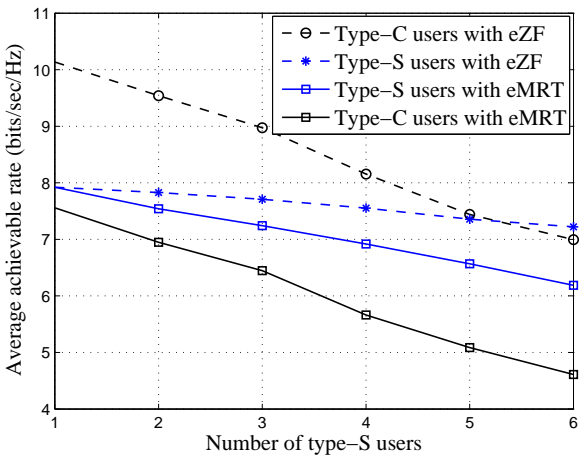}
\caption{ \textcolor{black}{Average achievable rate of type-C users versus numbers of type-S users with $ K=5$, $M=200$, $p_u$=10 dB, $T_{\mathrm{pilot}}=1$ OFDM symbol for uplink pilot transmission and MMSE receiver for type-C users.}}
\label{fig_S_num_rate}
\end{figure}

\begin{figure}[!h]
\centering
\includegraphics[width=3.5in]{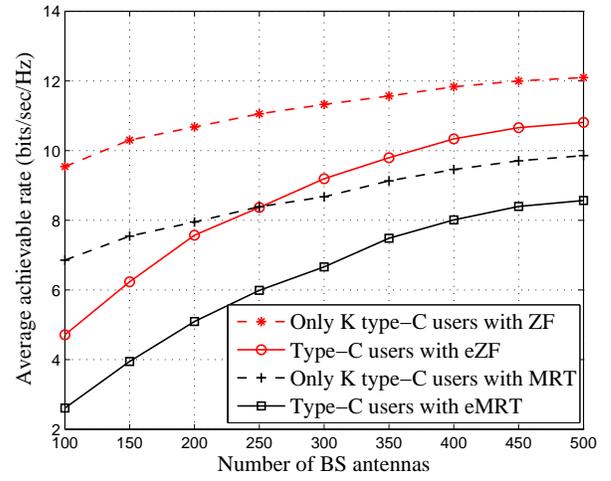}
\caption{ Average achievable rate for type-C users with $ K=5$, $ N=5$, \textcolor{black}{$p_u$=10 dB, $p_d$=10 dB, $T_{\mathrm{pilot}}=1$ OFDM symbol for uplink pilot transmission and MMSE receiver for type-C users.}}
\label{fig6 preamble}
\end{figure}

\textcolor{black}{However, due to the absence of type-S users' instantaneous CSI, the type-C users have to sacrifice useful channel information to eliminate the mutual interference between the type-C and type-S users.
To illustrate the impact of numbers of type-S users on the achievable rate of type-C users, Fig. \ref{fig_S_num_rate} is provided.
It is shown that the type-C users suffer more serious performance loss with the increasing number of type-S users.}
Nevertheless, Fig. \ref{fig6 preamble} illustrates that the performance loss of the type-C users will decrease with the increasing number of BS antennas.

In Fig. \ref{fig6 preamble}, the average achievable rate of type-C users is compared in two transmission modes.
\textcolor{black}{One mode is a massive MIMO system employing the conventional ZF and MRT methods and only serving $5$ type-C users.
The other is a massive MIMO system where the BS adopts the proposed precoding methods simultaneously serving $5$ type-C users and $5$ type-S users.}
The existence of the $5$ type-S users causes performance loss to the type-C users when compared to the conventional transmission mode.
However, the simulation result shows that the performance gap decreases with the increasing number of BS antennas.
Therefore, when the number of BS antennas is large enough, serving type-S users will not influence the performance of type-C users in massive MIMO systems.

\begin{figure}[!h]
\centering
\includegraphics[width=3.5in]{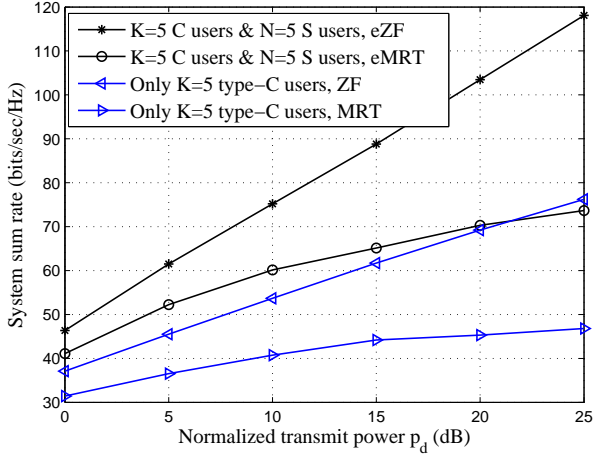}
\caption{ 
Sum rate comparison of the systems with limited pilot sequences when~\textcolor{black}{$M=200$,~$I= K=5$,~$N=5$, $p_u$=10 dB, $T_{\mathrm{pilot}}=1$ OFDM symbol for uplink pilot transmission and MMSE receiver for type-C users.}}
\label{fig7_preamble}
\end{figure}

\textcolor{black}{When pilot sequences are scarce in the systems and the length of pilot sequences is limited to $I$,
the BS could only schedule and serve $I$ type-C users with conventional precoding methods.
However, the BS is able to simultaneously serve $I$ type-C users and $N$ type-S users with the proposed methods, since the precoding design for the type-S users is only based on their statistical CSI and pilot-based channel estimation is avoided.}
\textcolor{black}{Thus, the sum rates of the systems employing the proposed eZF/eMRT methods and the conventional ZF/MRT methods are given as
\vspace{-0.3em}
\begin{align}\label{eq simu n1}
  R_{\mathrm{sum}}^{\mathrm{prop}} & =   \sum_{k=1}^{I}  {R}_{\mathrm{C},k} + \sum_{n=1}^{N} {R}_{\mathrm{S},n}  , \\
  R_{\mathrm{sum}}^{\mathrm{conv}} & =  \sum_{i=1}^{I}  {R}_i,
\end{align}
respectively.} Fig. \ref{fig7_preamble} explores the system sum rate comparison with~$I= K =5$.
\textcolor{black}{It is observed that even if the existence of type-S users leads to performance loss of type-C users, the system sum rate is significantly improved by $55.4 \%$  and $61.7 \%$ at~25 dB \textcolor{black}{compared to} the systems employing the conventional ZF and MRT methods, respectively.}
\textcolor{black}{Moreover, Fig.~7 is provided to explore how many BS antennas are needed for the proposed transmission
mode to outperform the conventional one. It is illustrated that the proposed transmission mode achieves higher system sum rate than the conventional MRT when the number of BS antennas is only larger than about 50. Besides, it outperforms the conventional ZF when the BS is only equipped with 79 antennas for the eZF precoding and 140 antennas for the eMRT precoding.
Moreover, the advantage of the proposed transmission mode becomes obvious with large scale of BS antennas.}

\begin{figure}[!h]
\centering
\includegraphics[width=3.5in]{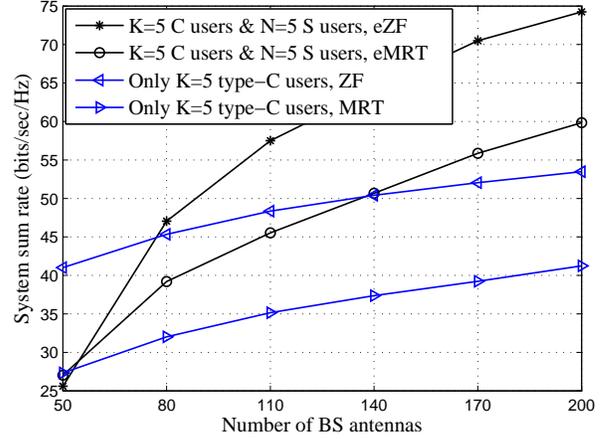}
\caption{ 
\textcolor{black}{Sum rate comparison of the systems versus different numbers of BS antennas when~$I= K=N=5$, $p_u$=10 dB, $p_d$=10 dB, $T_{\mathrm{pilot}}=1$ OFDM symbol for uplink pilot transmission and MMSE receiver for type-C~users.}}
\label{fig7-2_preamble}
\end{figure}

\begin{figure}[!h]
\centering
\includegraphics[width=3.5in]{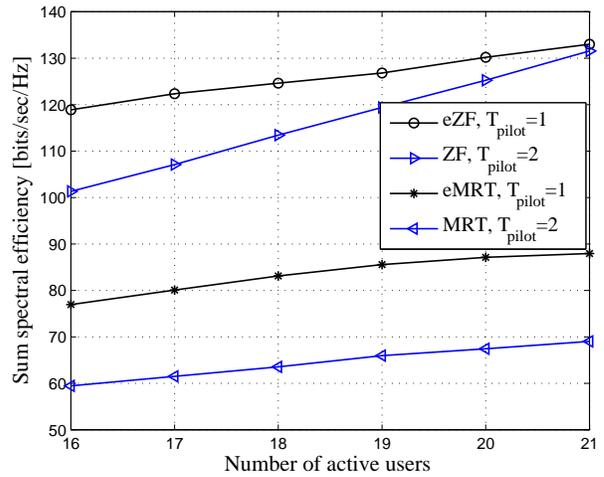}
\caption{ 
\textcolor{black}{Sum spectral efficiency versus numbers of active users with $M=300$, $p_u = 10$ dB, $p_d = 10 $ dB, $\theta_{\Delta}=5^{\circ}$, $T_{\mathrm{pilot}}$ OFDM symbols for uplink pilot transmission and MMSE receiver for type-C users.}}
\label{fig8_preamble}
\end{figure}

Moreover, pilot overhead\footnote{\textcolor{black}{Since statistical CSI is rather static and could be obtained by averaging over the past channel measurements. Therefore, we ignore the overhead for statistical CSI acquisition.}} \textcolor{black}{is also taken into account to evaluate the sum spectral efficiency (SE) of the systems employing the proposed eZF/eMRT methods and the conventional ZF/MRT methods. The sum SE is given as \cite{Marzetta} }
\vspace{-0.3em}
\begin{align}\label{eq simu n1}
  R_{\mathrm{SE}}^{\mathrm{prop}} & =\frac{ {T}_{\mathrm{slot}}- {T}_{\mathrm{overhead1}}}{ {T}_{\mathrm{slot}}} \left( \sum_{k=1}^{K}  {R}_{\mathrm{C},k} + \sum_{n=1}^{N} {R}_{\mathrm{S},n} \right), \\
  R_{\mathrm{SE}}^{\mathrm{conv}} & =\frac{ {T}_{\mathrm{slot}}- {T}_{\mathrm{overhead2}}}{ {T}_{\mathrm{slot}}} \sum_{i=1}^{K+N}  {R}_i,
\end{align}
\textcolor{black}{respectively, where $ {T}_{\mathrm{slot}}$ is the slot length, $ {T}_{\mathrm{overhead1}}$ and~$ {T}_{\mathrm{overhead2}}$ denote channel estimation and additional overhead\footnote{The additional overhead usually takes one OFDM symbol, including the overhead for uplink data transmission etc. Therefore, when there are~$T_{\mathrm{pilot}}$ OFDM symbols used for pilot transmission, the channel estimation efficiency is $\frac{ {T}_{\mathrm{slot}}- {T}_{\mathrm{overhead}}}{ {T}_{\mathrm{slot}}} = \frac{6-{T}_{\mathrm{pilot}}}{7}$.},} and $\frac{ {T}_{\mathrm{slot}}- {T}_{\mathrm{overhead}}}{ {T}_{\mathrm{slot}}}$ is the channel estimation efficiency, i.e., the ratio of time used for downlink data transmission to the total slot length.
Note that $ {T}_{\mathrm{overhead2}}$ is used for the channel estimation of $K+N$ users while $ {T}_{\mathrm{overhead1}}$ is only used for $K$ type-C users.
\textcolor{black}{
In Fig. \ref{fig8_preamble}, we compare the sum spectral efficiency $R_{\mathrm{SE}}^{\mathrm{prop}}$ and $R_{\mathrm{SE}}^{\mathrm{conv}}$.
To save the pilot overhead, the systems utilizing the proposed precoding methods tend to schedule $K=14$ users as type-C users and adopt $T_{\mathrm{pilot}}=1$ OFDM symbol for uplink pilot transmission.
Fig.~\ref{fig8_preamble} shows that
the systems with the proposed precoding methods use less pilot overhead and achieve higher sum SE.
Therefore, the proposed precoding methods are also good choices to save pilot overhead and improve spectral efficiency when the number of active users is large.}

\section{Conclusions}\label{section:conclusions}

\textcolor{black}{In this paper, the scenario where a BS simultaneously serves type-C and type-S users was considered.
The heuristic SBM method was first presented and we analytically demonstrated the impact of the mutual interference between type-C and type-S users on their achievable rate.
Then, the eZF and eMRT precoding methods were developed to remove the mutual interference.
When the proposed precoding methods are employed, the type-S users are able to obtain comparable average achievable rate with that of type-C users.
Moreover, the sum rate and sum SE of the systems are significantly enhanced with respect to the systems with conventional precoding methods.
Furthermore, thanks to the utilization of statistical CSI, pilot-based channel estimation for type-S users is avoided.
Then,
transmission delay is reduced and the BS is able to serve more active users with fixed pilot sequences.
Therefore, the new transmission mode with only imperfect instantaneous CSI of type-C users and statistical CSI  of type-S users is~feasible and suitable for prompt data transmission with the users without instantaneous CSI.}

\begin{appendices}
\appendices \numberwithin{equation}{section}

\section{\textcolor{black}{Useful Lemmas and Proofs}} \label{app_lemma}
\textcolor{black}{ \emph{Lemma 2 {(}\cite{Approxi}{)}:} Given any random variables $V_1$ and $V_2$, the approximation for the mean of a ratio $\frac{V_1}{V_2}$ is given as
\begin{equation} \label{heuheu2}
\hspace{-0.15em}\mathrm{E}\left\{  \frac{V_1}{V_2} \right\} \approx \frac{\mathrm{E}\left\{  V_1 \right\}}{\mathrm{E}\left\{  V_2 \right\}} -\frac{  \mathrm{Cov}\left(V_1,V_2 \right)  }{\mathrm{E}^2\left\{  V_2 \right\}} +\frac{  \mathrm{Var}\left( V_2 \right)  \mathrm{E}\left\{  V_1 \right\}  }{\mathrm{E}^3\left\{ V_2 \right\}}.
\end{equation}
}

\textcolor{black}{\emph{Lemma 3 (\cite[{Lemma 5}]{ShinWinJuly2008}):} Let $\mathbf{X}= \mathbf{\Psi}^{1/2} \mathbf{H} \mathbf{\Sigma}^{1/2} \sim \mathcal{CN}\left\{ \mathbf{0}_{M \times N}, \mathbf{\Psi}_{M \times M} , \mathbf{\Sigma}_{N \times N}\right\} $, where the elements of $\mathbf{H}$ are i.i.d. Then, for $\mathbf{A} \in \mathbb{C}^{M \times M} \geq 0$ and  $\mathbf{B} \in \mathbb{C}^{N \times N} \geq 0$, the~$k$th-order cumulant of $\mathrm{tr} \left(\mathbf{ AXBX}^H \right)$ is
\vspace{-0.3em}
\begin{equation}\label{eq 20170504.10}
\begin{split}
  &\mathrm{Cum}_k \left\{ \mathrm{tr} \left(\mathbf{ AXBX}^H \right) \right\} \\
 & \qquad= (k-1)! \mathrm{tr} \left\{  \left( \mathbf{A \Psi} \right)^k \right\} \mathrm{tr} \left\{  \left(\mathbf{ \Sigma} \mathbf{B} \right)^k \right\}.
\end{split}
\end{equation}}

\textcolor{black}{\emph{Lemma 4}: Let $\mathbf{Z} \triangleq \mathbf{h}\mathbf{h}^H$ with $\mathbf{h} \sim \mathcal{CN} \left( \mathbf{0}_{M \times 1}, \mathbf{I}_M  \right)$.
For~any matrices $\mathbf{A} \in \mathbb{C} ^{M \times M} $ and $\mathbf{B} \in \mathbb{C} ^{M \times M}$, we have
\begin{equation}\label{eq 20170504.1}
\begin{split}
& \mathbf{\mathbf{F}}\left(\mathbf{A},\mathbf{B}\right)  \triangleq \mathrm{ E} \left\{\mathrm{ tr}\left( \mathbf{ZA} \right) \mathrm{ tr}\left( \mathbf{ZB} \right) \right\}  \\
&= \mathrm{ tr}\left( \mathbf{A} \right) \mathrm{ tr}\left( \mathbf{B} \right)+\mathrm{ tr}\left( \mathbf{AB} \right)-\mathrm{ tr}\left( \mathbf{D}\left(\mathbf{A}\right) \mathbf{D}\left(\mathbf{B}\right) \right),
\end{split}
\end{equation}
where $\mathbf{D}\left( \mathbf{A} \right)$ denotes a diagonal matrix consisting of the diagonal elements of matrix $\mathbf{A}$.
\begin{proof}
Using the definition of trace operation, we have
\begin{equation}\label{eq 20170504.4}
\begin{split}
&\hspace{-1em}\mathrm{ E} \left\{\mathrm{ tr}\left( \mathbf{ZA} \right) \mathrm{ tr}\left( \mathbf{ZB} \right) \right\} \\
&\hspace{-1em}= \mathrm{E }\left\{\left(\sum_{i=1}^M \sum_{j=1}^M {Z}_{i,j} \mathbf{A}_{j,i}  \right)  \left( \sum_{m=1}^M \sum_{n=1}^M  {Z}_{m,n} \mathbf{B}_{n,m} \right)  \right\}\\
&\hspace{-1em}=\sum_{i=1}^M \sum_{j=1}^M \sum_{m=1}^M \sum_{n=1}^M \mathrm{E }\left\{  {Z}_{i,j}  {Z}_{m,n} \mathbf{A}_{j,i} \mathbf{B}_{n,m} \right\}\\
& \hspace{-1em} \overset{(a)}{=}\sum_{i=1}^M \sum_{m=1}^M \mathbf{A}_{i,i} \mathbf{B}_{m,m} +\sum_{i=1}^M \sum_{j=1,j \neq i }^M \mathbf{A}_{i,j} \mathbf{B}_{j,i} \\
&\hspace{-1em}= \mathrm{ tr}\left( \mathbf{A} \right) \mathrm{ tr}\left( \mathbf{B} \right) + \mathrm{ tr}\left( \mathbf{A}\mathbf{B} \right) -\mathrm{ tr}\left( \mathbf{D}\left(\mathbf{A}\right) \mathbf{D}\left(\mathbf{B}\right) \right),
\end{split}
\end{equation}
where $(a)$ is obtained by $Z_{i,j}=h_i h_j^H$ and $\mathbf{D}\left( \mathbf{A} \right)$ denotes a diagonal matrix consisting of the diagonal elements of~$\mathbf{A}$.
\end{proof}}

\section{\textcolor{black}{Proof of Equation (\ref{heuheu22})}} \label{app_rate_C}

\textcolor{black}{The proof is obtained by acquiring the three closed-form expressions of~(\ref{a1})--(\ref{a3}) as follows.}

\textcolor{black}{\emph{1)} Firstly, by taking (\ref{eq 20170429.3}) and~(\ref{heu c 1}) into (\ref{a1}), the average signal power~$S_{\mathrm{C},k}$ is given~as
\vspace{-0.5em}
\begin{align} \label{1.11}
S_{\mathrm{C},k} & = p_d \mathrm{E}\left\{ \hspace{-0.3em} \left|  \frac{ \mathbf{g}_{\mathrm{C},k}^H \widehat{\mathbf{g}}_{\mathrm{C},k} }{ \left\|\widehat{ \mathbf{g}}_{\mathrm{C},k} \right\|  }  \right|^2 \hspace{-0.3em} \right\}=\hspace{-0.3em}p_d  \mathrm{E}\left\{ \hspace{-0.3em} \left|  \left\| \widehat{\mathbf{g}}_{\mathrm{C},k} \right\|\hspace{-0.3em} +\hspace{-0.3em} \frac{ \widetilde{{ \mathbf{g}}}^H_{\mathrm{C},k} \widehat{ \mathbf{g}}_{\mathrm{C},k} }{ \left\| \widehat{\mathbf{g}}_{\mathrm{C},k} \right\|} \right|^2\hspace{-0.3em} \right\} \nonumber \\
& \overset{(a)}{=} p_d \mathrm{E }\left\{ \left\| \widehat{\mathbf{g}}_{\mathrm{C},k} \right\|^2 \right\}+ p_d \mathrm{E }\left\{\frac{\left| \widetilde{{ \mathbf{g}}}^H_{\mathrm{C},k} \widehat{ \mathbf{g}}_{\mathrm{C},k}\right|^2 }{ \left\| \widehat{\mathbf{g}}_{\mathrm{C},k} \right\|^2}  \right\},
\end{align}
where $(a)$ is due to the independence of $\widehat{ \mathbf{g}}_{\mathrm{C},k} $ and $\widetilde{{ \mathbf{g}}}^H_{\mathrm{C},k} $.
Set~$V_3 \triangleq \left| \widetilde{{ \mathbf{g}}}_{\mathrm{C},k} \widehat{ \mathbf{g}}_{\mathrm{C},k}\right|^2$ and $V_4 \triangleq  \left\| \widehat{\mathbf{g}}_{\mathrm{C},k} \right\|^2$.
Based on \emph{Lemma}~2, the second item in (\ref{1.11}) is further derived~as
\begin{align}
 & \mathrm{E }\left\{\frac{\left| \widetilde{{ \mathbf{g}}}^H_{\mathrm{C},k} \widehat{ \mathbf{g}}_{\mathrm{C},k}\right|^2 }{ \left\| \widehat{\mathbf{g}}_{\mathrm{C},k} \right\|^2}  \right\} =\mathrm{E }\left\{\frac{V_3 }{V_4}  \right\} \nonumber \\
 & \approx \frac{\mathrm{E}\left\{  V_3 \right\}}{\mathrm{E}\left\{  V_4 \right\}} -\frac{  \mathrm{Cov}\left(V_3,V_4 \right)  }{\mathrm{E}^2\left\{  V_4 \right\}} +\frac{  \mathrm{Var}\left( V_4 \right)  \mathrm{E}\left\{  V_3 \right\}  }{\mathrm{E}^3\left\{ V_4 \right\}}.\label{eq 20170504.5}
\end{align}
Since $\widetilde{{ \mathbf{g}}}^H_{\mathrm{C},k}$ and $\widehat{ \mathbf{g}}_{\mathrm{C},k}$ are statistically independent, $\mathrm{E}\left\{  V_3 \right\}$ and~$\mathrm{E}\left\{  V_4 \right\}$ are given as
\begin{align}\label{eq 20170504.6}
 \mathrm{E}\left\{  V_3\right\} & =\mathrm{tr}\left( \mathbf{\Delta}_{\mathrm{C},k}  \widehat{\mathbf{ \Phi}}_{{\mathrm{C},k}}  \right),\\
 \mathrm{E}\left\{  V_4 \right\} & =\mathrm{tr}\left( \widehat{\mathbf{ \Phi}}_{{\mathrm{C},k}}  \right),
\end{align}
respectively. Moreover, we have
\begin{align}\label{eq 20170504.9}
\mathrm{E }\left\{ V_3 V_4 \right\}  & = \mathrm{E }_{\widetilde{\mathbf{g}}_{\mathrm{C},k},\widehat{\mathbf{g}}_{\mathrm{C},k}} \left\{ \left|  \widetilde{\mathbf{g}}^H_{\mathrm{C},k}\widehat{\mathbf{g}}_{\mathrm{C},k} \right|^2 \left\|  \widehat{\mathbf{g}}_{\mathrm{C},k} \right\|^2\right\} \nonumber \\
& = \mathrm{E }_{\widetilde{\mathbf{g}}_{\mathrm{C},k},\widehat{\mathbf{g}}_{\mathrm{C},k}} \left\{ \widetilde{\mathbf{g}}^H_{\mathrm{C},k} \widehat{\mathbf{g}}_{\mathrm{C},k}\widehat{\mathbf{g}}^H_{\mathrm{C},k} \widetilde{\mathbf{g}}_{\mathrm{C},k} \widehat{\mathbf{g}}^H_{\mathrm{C},k} \widehat{\mathbf{g}}_{\mathrm{C},k} \right\} \nonumber \\
& \overset{(a)}{=} \mathrm{E }_{\widetilde{\mathbf{g}}_{\mathrm{C},k},\widehat{\mathbf{g}}_{\mathrm{C},k}} \left\{ \widetilde{\mathbf{g}}^H_{\mathrm{C},k} \mathbf{M}_1 \widetilde{\mathbf{g}}_{\mathrm{C},k} \mathbf{M}_2 \right\}\nonumber \\
& = \mathrm{E }_{\widetilde{\mathbf{g}}_{\mathrm{C},k}} \left\{ \mathrm{tr }\left\{  \mathbf{M}_1 \widetilde{ \mathbf{g}} _{\mathrm{C},k}\mathbf{M}_2 \widetilde{\mathbf{g}}^H_{\mathrm{C},k}  \right\}  \right\}\nonumber \\
& \overset{(b)}{=} \mathrm{E }_{\mathbf{M}_1 ,\mathbf{M}_2} \left\{ \mathrm{tr} \left( \mathbf{M}_1\mathbf{\Delta}_{\mathrm{C},k} \right)  \mathrm{tr} \left( \mathbf{M}_2 \right)  \right\},
\end{align}
where $(a)$ is obtained by setting $\mathbf{M}_1 \triangleq \widehat{\mathbf{g}}_{\mathrm{C},k}\widehat{\mathbf{g}}^H_{\mathrm{C},k} $ and $\mathbf{M}_2 \triangleq \widehat{\mathbf{g}}^H_{\mathrm{C},k}\widehat{\mathbf{g}} _{\mathrm{C},k} $ and $(b)$ follows from \emph{Lemma} 3 with $k=1$, $\Psi = \mathbf{\Delta}_{\mathrm{C},k}$ and $\mathbf{\Sigma}=1$.
Substituting~$\mathbf{M}_1$, $\mathbf{M}_2$ and $ \widehat{\mathbf{g}}_{\mathrm{C},k} ={\widehat{\mathbf{ \Phi}}}^{1/2}_{\mathrm{C},k} \widehat{\mathbf{h}}_{\mathrm{C},k}  $ with $\widehat{\mathbf{h}}_{\mathrm{C},k} \sim \mathcal{CN }  \left( \mathbf{0}_{1 \times M}, \mathbf{I}_M \right)$ into (\ref{eq 20170504.9}), we have
\begin{align}\label{eq 20170504.11}
&\mathrm{E }\left\{ V_3 V_4 \right\} \nonumber \\
 &= \mathrm{E }\left\{ \mathrm{tr} \left(\widehat{\mathbf{h}}_{\mathrm{C},k} \widehat{\mathbf{h}}_{\mathrm{C},k}^H \widehat{\mathbf{ \Phi}}_{\mathrm{C},k}^{1/2} \mathbf{\Delta}_{\mathrm{C},k}  \widehat{\mathbf{ \Phi}}_{\mathrm{C},k}^{1/2}   \right) \mathrm{tr} \left(\widehat{\mathbf{h}}_{\mathrm{C},k} \widehat{\mathbf{h}}_{\mathrm{C},k}^H  \widehat{\mathbf{ \Phi}}_{\mathrm{C},k} \right)  \right\} \nonumber \\
 & \overset{(a)}{=} \mathbf{\mathbf{F}} \left(\widehat{\mathbf{ \Phi}}_{\mathrm{C},k}^{1/2} \mathbf{\Delta}_{\mathrm{C},k}  \widehat{\mathbf{ \Phi}}_{\mathrm{C},k}^{1/2},  \widehat{\mathbf{ \Phi}}_{\mathrm{C},k} \right),
\end{align}
where $(a)$ results from \emph{Lemma}~4.}

\textcolor{black}{Furthermore, $\mathrm{Var}\left( V_4 \right)$ in (\ref{eq 20170504.5}) could be further written~as
\begin{equation}\label{eq 20170520.1}
\begin{split}
 \mathrm{Var}\left( V_4 \right)
 & = \mathrm{E}\left\{  V_4^2  \right\}-\mathrm{E}^2\left\{  V_4 \right\}\\
 & = \mathrm{E}\left\{ \left\| \widehat{\mathbf{g}}_{\mathrm{C},k} \right\|^4  \right\}-\mathrm{tr}^2 \left( \widehat{\mathbf{ \Phi}}_{\mathrm{C},k}  \right).
\end{split}
\end{equation}
The first item $\mathrm{E}\left\{ \left\| \widehat{\mathbf{g}}_{\mathrm{C},k} \right\|^4  \right\}$ is given~as
\begin{equation}\label{eq 20170520.2}
\begin{split}
&\mathrm{E}\left\{ \left\| \widehat{\mathbf{g}}_{\mathrm{C},k} \right\|^4  \right\} = \mathrm{E}\left\{  \widehat{\mathbf{g}}_{\mathrm{C},k}^H \widehat{\mathbf{g}}_{\mathrm{C},k} \widehat{\mathbf{g}}_{\mathrm{C},k}^H \widehat{\mathbf{g}}_{\mathrm{C},k} \right\}\\
&= \mathrm{E}\left\{  \widehat{\mathbf{h}}_{\mathrm{C},k}^H {\widehat{\mathbf{ \Phi}}}_{\mathrm{C},k}  \widehat{\mathbf{h}}_{\mathrm{C},k} \widehat{\mathbf{h}}_{\mathrm{C},k}^H {\widehat{\mathbf{ \Phi}}}_{\mathrm{C},k} \widehat{\mathbf{h}}_{\mathrm{C},k} \right\}\\
&= \mathrm{E}\left\{ \mathrm{tr}\left( \widehat{\mathbf{h}}_{\mathrm{C},k}^H {\widehat{\mathbf{ \Phi}}}_{\mathrm{C},k}  \widehat{\mathbf{h}}_{\mathrm{C},k}\right) \mathrm{tr}\left(  \widehat{\mathbf{h}}_{\mathrm{C},k}^H {\widehat{\mathbf{ \Phi}}}_{\mathrm{C},k} \widehat{\mathbf{h}}_{\mathrm{C},k}\right) \right\} \\
& = \mathrm{E}\left\{ \mathrm{tr}\left( \widehat{\mathbf{h}}_{\mathrm{C},k} \widehat{\mathbf{h}}_{\mathrm{C},k}^H {\widehat{\mathbf{ \Phi}}}_{\mathrm{C},k}  \right) \mathrm{tr}\left(  \widehat{\mathbf{h}}_{\mathrm{C},k} \widehat{\mathbf{h}}_{\mathrm{C},k}^H {\widehat{\mathbf{ \Phi}}}_{\mathrm{C},k}\right) \right\} \\
& \overset{(a)}{=} \mathbf{F} \left( {\widehat{\mathbf{ \Phi}}}_{\mathrm{C},k}, {\widehat{\mathbf{ \Phi}}}_{\mathrm{C},k} \right),
\end{split}
\end{equation}
where $(a)$ follows from \emph{Lemma}~4.
Thus, combining~(\ref{eq 20170504.5})--(\ref{eq 20170520.2}) with (\ref{1.11}), we obtain the average signal power~$S_{\mathrm{C},k}$~as
\vspace{-0.3em}
\begin{equation}\label{eq 20170504.12}
\begin{split}
&S_{\mathrm{C},k}= p_d \Bigg(\mathrm{tr} \left(\widehat{\mathbf{ \Phi}}_{\mathrm{C},k}  \right)+  \frac{\mathrm{tr} \left(\widehat{\mathbf{ \Phi}}_{\mathrm{C},k} \mathbf{\Delta}_{\mathrm{C},k} \right)}{\mathrm{tr} \left(\widehat{\mathbf{ \Phi}}_{\mathrm{C},k}  \right)} \\
&- \frac{\mathbf{\mathbf{F}}\left( \widehat{\mathbf{ \Phi}}_{\mathrm{C},k}^{1/2} \mathbf{\Delta}_{\mathrm{C},k} \widehat{\mathbf{ \Phi}}_{\mathrm{C},k}^{1/2} , \widehat{\mathbf{ \Phi}}_{\mathrm{C},k} \right)-\mathrm{tr}\left(\widehat{\mathbf{ \Phi}}_{\mathrm{C},k} \mathbf{\Delta}_{\mathrm{C},k} \right) \mathrm{tr} \left(\widehat{\mathbf{ \Phi}}_{\mathrm{C},k}  \right) }{\mathrm{tr} ^2\left(\widehat{\mathbf{ \Phi}}_{\mathrm{C},k}  \right)}\\
&+\frac{ \left[ \mathbf{ F}\left( \widehat{\mathbf{ \Phi}}_{\mathrm{C},k},\widehat{\mathbf{ \Phi}}_{\mathrm{C},k} \right)  - \mathrm{tr} ^2\left(\widehat{\mathbf{ \Phi}}_{\mathrm{C},k}  \right) \right] \mathrm{tr}\left(\widehat{\mathbf{ \Phi}}_{\mathrm{C},k} \mathbf{\Delta}_{\mathrm{C},k} \right) }{\mathrm{tr} ^3\left(\widehat{\mathbf{ \Phi}}_{\mathrm{C},k}  \right)}  \Bigg)\\
& =p_d \Bigg(\mathrm{tr} \left(\widehat{\mathbf{ \Phi}}_{\mathrm{C},k}  \right)\hspace{-0.3em} +\hspace{-0.3em} \frac{\mathrm{tr} \hspace{-0.3em}\left(\hspace{-0.3em}\widehat{\mathbf{ \Phi}}_{\mathrm{C},k} \mathbf{\Delta}_{\mathrm{C},k} \hspace{-0.3em}\right)}{\mathrm{tr} \hspace{-0.3em} \left(\widehat{\mathbf{ \Phi}}_{\mathrm{C},k}  \right)}\hspace{-0.3em} - \hspace{-0.3em}  \frac{ \mathbf{\mathbf{F}}\hspace{-0.3em}\left( \hspace{-0.3em}\widehat{\mathbf{ \Phi}}_{\mathrm{C},k}^{1/2} \mathbf{\Delta}_{\mathrm{C},k} \widehat{\mathbf{ \Phi}}_{\mathrm{C},k}^{1/2} , \widehat{\mathbf{ \Phi}}_{\mathrm{C},k} \hspace{-0.3em}\right)\hspace{-0.3em}   }{\mathrm{tr} ^2\left(\widehat{\mathbf{ \Phi}}_{\mathrm{C},k}  \right)} \\
 &+ \frac{ \mathbf{ F}\hspace{-0.3em} \left( \widehat{\mathbf{ \Phi}}_{\mathrm{C},k},\widehat{\mathbf{ \Phi}}_{\mathrm{C},k}\hspace{-0.3em} \right)    \mathrm{tr} \left(\widehat{\mathbf{ \Phi}}_{\mathrm{C},k} \mathbf{\Delta}_{\mathrm{C},k} \right)}{\mathrm{tr} ^3\left(\widehat{\mathbf{ \Phi}}_{\mathrm{C},k}  \right)} \Bigg).
\end{split}
\end{equation}}

\textcolor{black}{\emph{2)}
By substituting (\ref{heu c 1}) into (\ref{a2}), the average interference power $I_{\mathrm{C1},k}$ is obtained as
\begin{equation}\label{eq 20170504.14}
\begin{split}
 & I_{\mathrm{C1},k}  = p_d  \sum\limits_{i = 1,i \ne k}^K  \mathrm{E }\left\{\frac{\left| {{ \mathbf{g}}}^H_{\mathrm{C},k} \widehat{ \mathbf{g}}_{\mathrm{C},i}\right|^2 }{ \left\| \widehat{\mathbf{g}}_{\mathrm{C},i} \right\|^2}  \right\}.
\end{split}
\end{equation}}
\textcolor{black}{With a similar derivation of (\ref{eq 20170504.5}), $ I_{\mathrm{C1},k}$ is given as
\begin{equation}\label{eq 20170521.1}
\begin{split}
  &I_{\mathrm{C1},k} =p_d \sum\limits_{i = 1,i \ne k}^K   \Bigg ( \frac{\mathrm{tr} \left(\widehat{\mathbf{ \Phi}}_{\mathrm{C},i} \mathbf{\Phi}_{\mathrm{C},k} \right)}{\mathrm{tr} \left(\widehat{\mathbf{ \Phi}}_{\mathrm{C},i}  \right)} \\
  &-\hspace{-0.3em} \frac{ \mathbf{\mathbf{F}}\left( \widehat{\mathbf{ \Phi}}_{\mathrm{C},k}^{1/2} \mathbf{\Phi}_{\mathrm{C},k} \widehat{\mathbf{ \Phi}}_{\mathrm{C},i}^{1/2} , \widehat{\mathbf{ \Phi}}_{\mathrm{C},i} \right)   }{\mathrm{tr} ^2\left(\widehat{\mathbf{ \Phi}}_{\mathrm{C},i}  \right)}  \hspace{-0.3em}  +\hspace{-0.3em} \frac{ \mathbf{ F}\hspace{-0.3em} \left( \widehat{\mathbf{ \Phi}}_{\mathrm{C},i},\widehat{\mathbf{ \Phi}}_{\mathrm{C},i} \right)    \hspace{-0.3em}  \mathrm{tr} \left(\widehat{\mathbf{ \Phi}}_{\mathrm{C},i} \mathbf{\Phi}_{\mathrm{C},k} \right)}{\mathrm{tr} ^3\left(\widehat{\mathbf{ \Phi}}_{\mathrm{C},i}  \right)}\hspace{-0.3em}  \Bigg).
\end{split}
\end{equation}}

\textcolor{black}{\emph{3)} Substituting (\ref{heuheu9}) into (\ref{a3}), we obtain the interference power $I_{\mathrm{C2},k} $ as
\vspace{-0.5em}
\begin{equation}\label{add proof1}
\begin{split}
\hspace{-1.0em}I_{\mathrm{C2},k} & = p_d  \sum\limits_{n = 1}^N  \hspace{-0.3em} \mathbf{u}^H_\mathrm{max} \left ( \mathbf{\Phi}_{\mathrm{S},n} \right) \mathrm{ E } \left\{ \mathbf{g}_{\mathrm{C},k} \mathbf{g}_{\mathrm{C},k}^H \right \} \mathbf{u}_\mathrm{max} \left ( \mathbf{\Phi}_{\mathrm{S},n} \right)\\
& =p_d  \sum\limits_{n = 1}^N  \mathbf{u}^H_\mathrm{max} \left ( \mathbf{\Phi}_{\mathrm{S},n} \right)  \mathbf{\Phi} _{{\mathrm{C},k}}  \mathbf{u}_\mathrm{max} \left ( \mathbf{\Phi}_{\mathrm{S},n} \right).
\end{split}
\end{equation}}

\textcolor{black}{Combining (\ref{eq 20170504.12})--(\ref{add proof1}) with (\ref{1.11}), we complete the proof.}

\end{appendices}

\end{document}